 \RequirePackage[2020-02-02]{latexrelease}
\documentclass[floatfix,aps,prl,twocolumn,groupedaddress,showpacs,english]{revtex4}
\usepackage{comment}
\usepackage{color,soul}
\usepackage{placeins}
\usepackage[T1]{fontenc}
\usepackage[latin9]{inputenc}
\usepackage{amsmath}
\usepackage{amssymb}
\usepackage{graphicx}
\usepackage{subfig}
\usepackage{xcolor}

\usepackage{caption}
\usepackage{appendix}
\usepackage{wasysym}
\usepackage[version=4]{mhchem}
\usepackage{collcell}

\usepackage{comment}
\usepackage{babel}
\usepackage{chngcntr}
\begin{document}
\title{Excitation of absorbing exceptional points in the time domain}
\author{Asaf Farhi\textsuperscript{1} , Ahmed Mekawy\textsuperscript{3,4} , Andrea Al\`u\textsuperscript{3,4,5}, and Douglas Stone\textsuperscript{1,2}}
\affiliation{\textsuperscript{1}Department of Applied Physics, Yale University, New Haven, Connecticut
06520, USA}\vspace*{2cm}
\affiliation{\textsuperscript{2}Yale Quantum Institute, Yale University, New Haven, Connecticut 06520,
USA}
\affiliation{\textsuperscript{3} Photonics Initiative, Advanced Science Research Center, City University
of New York, New York, New York 10031, USA}
\affiliation{\textsuperscript{4}Department of Electrical Engineering, City College of The City University
of New York, New York, New York 10031, }
\affiliation{\textsuperscript{5}Physics Program, Graduate Center, City University of New York, New
York, New York 10016, USA}

\begin{abstract}

We analyze the time-domain dynamics of resonators supporting exceptional points (EPs), at which both the eigenfrequencies and the eigenmodes associated with perfect capture of an input wave coalesce.  We find that a time-domain signature of the EP is an expansion of the class of waveforms which can be perfectly captured. We show that such resonators have improved performance for storage or transduction of energy. They also can be used to convert between waveforms within this class. We analytically derive these features and demonstrate them for several examples of coupled optical resonator systems. 


\end{abstract}

\maketitle
\clearpage
\newpage
\pagebreak
\pagebreak
\newpage
\newpage
\pagebreak
\pagebreak
\newpage

Optimal energy transfer via electromagnetic waves is of high importance for a variety of applications, such as transmitting information in integrated photonics and quantum processors, and energy transduction in ablation and solar cells. The limit of perfect transfer (zero scattering) in general has solutions only at discrete (possibly complex) frequencies. 
One form of optimal transduction of an electromagnetic wave is Coherent Perfect Absorption (CPA), the time-reversed process of lasing at threshold, in which by tuning the degree of absorption in a structure, a specific continuous wave (CW) input at real frequency will be perfectly absorbed   \mbox{\cite{chong2010coherent}}.  It has recently been shown that an analogous phenomenon can be achieved in lossless systems by exciting a zero scattering state at a complex frequency with an exponentially rising input wave. In this case, the system will simply store the input until the ramp is turned off and the energy is released \mbox{\cite{baranov2017coherent}}. Showing that it is possible to access such states  raises a natural question: what are the time domain signatures of \emph{degeneracies} of these states, known as exceptional points (EPs). 

 When a non-Hermitian system is tuned to have a degeneracy, two or more eigenvalues \emph{and} eigenfunctions coalesce. EPs of resonances have been shown to lead to enhanced wave-matter interactions, improved sensing, asymmetric state transfer, and novel lasing behavior \mbox{\cite{el2007theory,makris2008beam,wiersig2011structure,wiersig2014chiral,kato2013perturbation,moiseyev2011non,bender1998real,guo2009observation,ruter2010observation,heiss2012physics,doppler2016dynamically,xu2016topological,feng2017non,zhang2018hybrid,pick2017general,zhang2019quantum,zhen2015spawning,chen2020revealing,wang2020electromagnetically,naghiloo2019quantum,song2021plasmonic,miri2019exceptional}}. 
Recently, phenomena associated with the presence of EPs of CPAs have been studied and probed in the frequency domain, focusing on real frequencies. An anomalous quartic line broadening was predicted due to the presence of the EP \mbox{\cite{sweeney2019perfectly}}, and observed in a coupled ring resonator system, \mbox{\cite{wang2021coherent}}.
However, this earlier work considered only the CW input.

Here we show that the implication of this coalescence of eigenfunctions at a real or virtual CPA EP is that a second wave-equation solution arises, of the form $(vt-z)e^{ikz-i\omega t}$, where $v$ is the propagation speed and $k$ is the wavevector. Here $\omega, 
k \equiv (\omega/v)$ are real for CPA EP and complex for virtual CPA EP. More generally for an $m^{th}$ order degeneracy, solutions of the form $(vt-z)^{m-1}e^{ikz -i\omega t}$ and all lower powers exist and any superposition thereof satisfies the zero scattering boundary condition. 
These modes are growing temporally and decaying spatially along the propagation axis. Importantly, from time-reversal arguments one can show that the time reversal of these more general waveforms describes the emission of systems at resonance EPs \mbox{\cite{heugel2010analogy,wenner2014catching}}. The possibility of exciting a zero scattering state with any superposition of these waveforms hasn't been explored prior to this work. Not only is this of fundamental interest, but it allows increased flexibility in exciting such a structure without
generating reflections; we will show that this leads to improved impedance matching of finite pulses and the ability to load and potentially empty a cavity faster. 

The fact that these waveforms are reflectionless can be seen from the following general argument, which applies both to CPA and the recently discussed Reflectionless Scattering Modes (RSMs). While at a CPA, with appropriate spatial excitation, there is no reflection to any of the input channels, RSMs are states which are defined by zero reflection into a chosen subset of the input channels \cite{sweeney2020theory,dhia2018trapped}, but for which the input is partially or fully transmitted into the complementary outgoing channels. For both CPA and RSMs, an eigenvalue of a suitably defined reflection matrix is zero at a particular $\omega$, and an EP of order $m$ is created by the coalescence of $m$ such frequencies and eigenvectors at $\omega_{\mathrm{EP}}$. The single remaining reflection eigenvalue vanishes as $\rho (\omega) \sim(\omega -\omega_{\mathrm{EP}})^m$ at the EP \cite{sweeney2019perfectly}, and hence all derivatives of $\rho (\omega)$ up to order $m-1$ vanish. Specializing this to the case $m=2$, which will be our focus here, we consider inputs at $z=0$ of $e^{i\omega_{1}t}$ and $te^{i\omega_{1}t}$; we 
Fourier transform (FT) them, and then apply the inverse FT to their product with $\rho(\omega)$:
\begin{align}
&\mathcal{F}\left(e^{i\omega_{1}t}\right)=\delta\left(\omega-\omega_{1}\right),~\mathcal{F}\left(te^{i\omega_{1}t}\right)=\delta'\left(\omega-\omega_{1}\right),\nonumber\\
&\int\delta\left(\omega-\omega_{1}\right)\rho\left(\omega\right)e^{i\omega t}d\omega=\rho\left(\omega_{1}\right)e^{i\omega_{1}t},\\
&\int\delta'\left(\omega-\omega_{1}\right)\rho\left(\omega\right)e^{i\omega t}d\omega=\rho'\left(\omega_{1}\right)e^{i\omega_{1}t}+\rho\left(\omega_{1}\right)ite^{i\omega_{1}t}\nonumber.
\end{align} 
 Thus, at a CPA or RSM EP, since $\rho\left(\omega\right)=\rho'\left(\omega\right)=0$, there is no reflection of
any linear combination of such inputs at real or complex $\omega_{\mathrm{EP}}$. This obviously applies to the inputs  $t^{m-1}\exp(i\omega_1 t)$  with $m\geq 3$ at higher-order EPs. To satisfy the wave equation, the input has to be of the form $f(vt-z)$ and we obtain the explicit solution mentioned above.
In addition, it can be seen that at a generic CPA, in which only $\rho(\omega)=0$, the growing input $t\exp(i\omega t)$ is reflected but is converted to the constant-amplitude output $\exp(i\omega t).$ We will discuss such conversion processes, which generalize to higher orders, briefly below and in the SM Sec. 1.


The above argument neglects the effect of the turn on and off of the input wave. To estimate the effect of the turn on at a CPA and CPA EP, we FT the inputs of the form $\tilde{\theta}(t)e^{i\omega_1 t}$
and $\tilde{\theta}(t)te^{i\omega_1 t},$ where $\tilde{\theta}(t)\equiv \theta(t)\,\mathrm{or}\, (\tanh(t)+1)/2$  and multiply the results by $\rho=r$, the reflection coefficient, to get
\begin{align}&\lim_{\omega\rightarrow \omega_{\mathrm{CPA}}}\mathcal{F}\left(\tilde{\theta}\left(t\right)e^{i\omega t}\right)\cdot r_{\mathrm{CPA}}  \propto\frac {\omega-\omega_{\mathrm{CPA}}}{\omega-\omega_{\mathrm{CPA}}}=\mathrm{constant},\nonumber\\
&\lim_{\omega\rightarrow \omega_{\mathrm{EP}}}\mathcal{F}\left(\tilde{\theta}\left(t\right)e^{i\omega t}\right)\cdot r_{\mathrm{CPA\,EP}}  \propto \frac {\left(\omega-\omega_{\mathrm{EP}}\right)^2}{\omega-\omega_{\mathrm{EP}}}=\left(\omega-\omega_{\mathrm{EP}}\right),\nonumber\\
&\lim_{\omega\rightarrow \omega_{\mathrm{EP}}}\mathcal{F}\left(\tilde{\theta}\left(t\right)te^{i\omega t}\right)\cdot r_{\mathrm{CPA\,EP}} \propto\frac {\left(\omega-\omega_{\mathrm{EP}}\right)^2}{\left(\omega-\omega_{\mathrm{EP}}\right)^2} =\mathrm{constant.}\nonumber
\end{align}
Interestingly, we see that for input $e^{i\omega t}$ the effect of the  turn on (introducing a pole) is strongly damped at CPA EP, as the response at $\omega_{\mathrm{EP}}$ is zero, whereas at a CPA it is a constant, corresponding to a slower temporal decay for the same input. 
 In addition, while the response to $te^{i\omega t}$ at a CPA EP is also constant, the input in this second case is much larger for long excitation times, which implies smaller relative reflection. 
Moreover, due to the linear dependency on $t$ in $t\exp(i\omega t)$, the scattered field that originates from the incoming field 
at earlier times is smaller in magnitude than the current input and the scattered field before the destructive interference starts is small, which results in small relative reflection compared to $\exp(i\omega t)$. 

We now demonstrate these general properties in an analytically solvable model oriented towards optics; similar excitation properties apply in AC circuits, acoustics, and quantum scattering. For simplicity we take a structure with a single input channel, terminated with a perfect mirror. 
\begin{figure}[ht]
\subfloat[]{
\includegraphics[width=7.5cm]{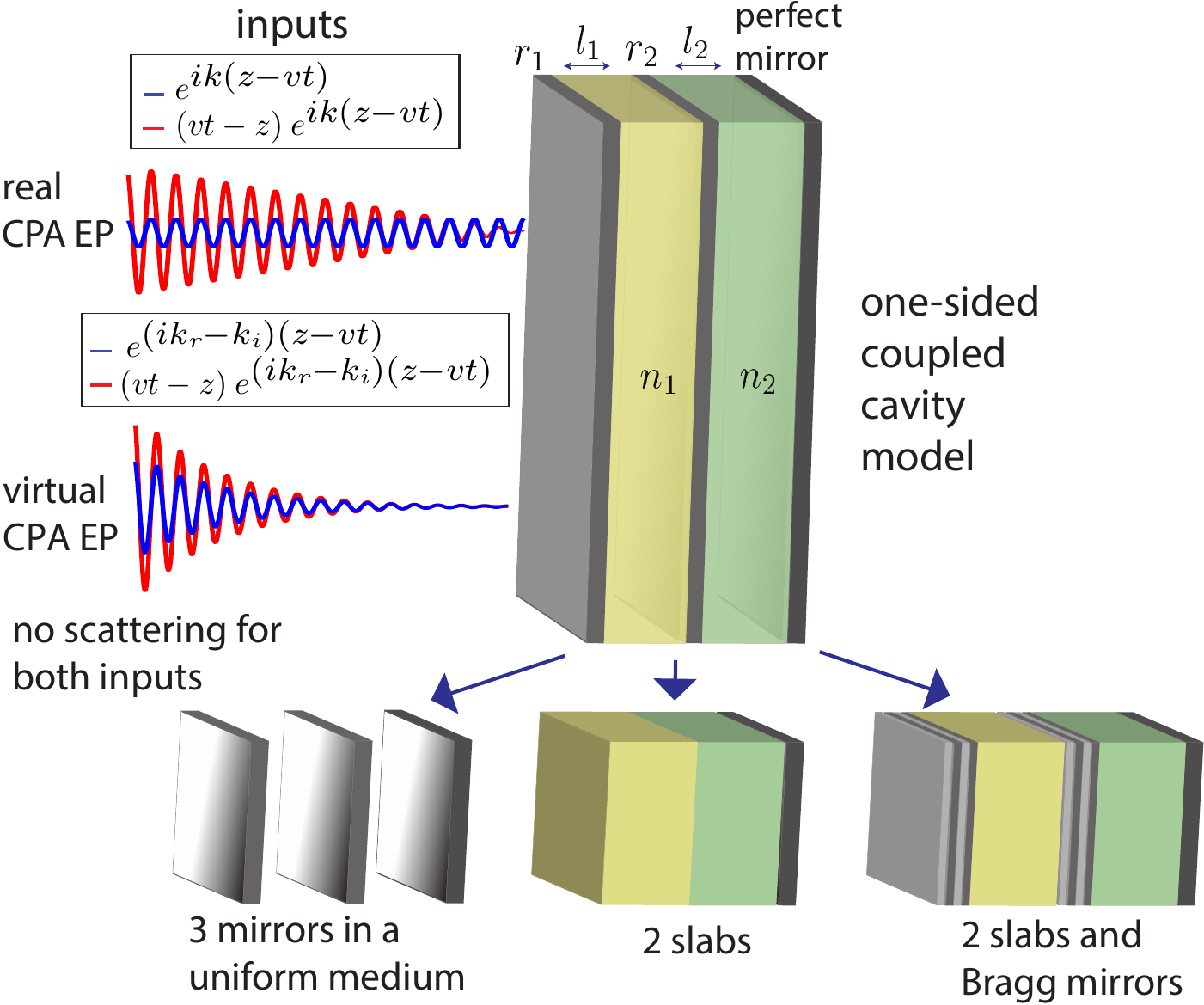}}\\
\subfloat[]{
 \includegraphics[width=3.6cm]{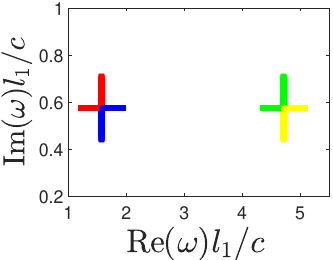}
 }\subfloat[]{
\includegraphics[width=4.2cm]{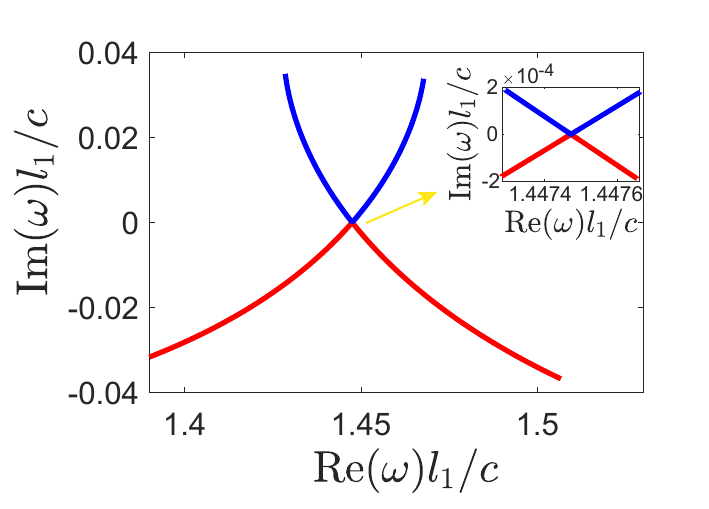}}
\caption{(a) Model of two coupled cavities terminated on the right with a perfect mirror and separated by two lossless partially reflecting mirrors of 
reflectivities $r_{1},r_{2}.$ 
At a real CPA EP $\omega_{\mathrm{EP}}$ is real and there is no scattering of the inputs associated with $\exp\left(i\omega t\right)$ and $t\exp\left(i\omega t\right).$  At a
virtual CPA EP $\omega_{\mathrm{EP}}$ is complex and there is no scattering of
the inputs associated with $\exp\left(i\omega_{r}t+\Gamma t\right)$ and $t\exp\left(i\omega_{r}t+\Gamma t\right),$
where $\omega_{r}\equiv\mathrm{Re}\left(\omega\right)=k_rv$ and $\Gamma\equiv\mathrm{Im}\left(\omega\right)=k_iv.$ This model can represent 3 mirrors in a uniform medium, 2 slabs, and 2 slabs with Bragg mirrors. (b) Example of a virtual CPA for the two-slab setup with $r_{2\,\mathrm{EP}}=0.574 \,\,(n_2=4.52),  
r_{1\mathrm{EP}}=0.1\, (n_1=1.22),l_{1}=l_{2}=1$ where we varied $r_{2}.$ For this special case, additional CPA EPs occur in each free spectral range. \label{fig:VitrualCPAEP} (c) Meeting of two CPAs at a real CPA EP for the two-slab and Bragg-mirror setup as we varied $l_{2}.$ The EP is at $N_{1}=5\,\,\,N_{2}=7,\,\,\,n_{3}=1.9,$ $n_{4}=1.5 , n_{1}=1.083+0.005i,\,\,\,n_{2}=2.17+0.107i ,\,\,\,l_{1}=1,\,\,\,l_{2\,\mathrm{EP}}=1.5.$\label{fig:Meeting-of-two_Bragg}}
\end{figure}
A property of the response at an EP, which has been exploited for sensing applications, is that the EP leads to higher sensitivity of the eigenvalues to perturbations in the parameters of the system \cite{chen2017exceptional,park2020symmetry,lai2019observation,wu2019observation}.  In the context of modeling  this implies a higher degree of difficulty in locating EPs by pure numerical search. To avoid this, as a first step we consider the single-port three-mirror model shown in Fig. 1, which is analytically tractable and still rather general.  It consists of two regions of length $l_1,l_2$ and uniform refractive index $n_1,n_2$ terminated by a perfect mirror. There are lossless mirrors of reflectivity $r_1,r_2$ on the left surface of each region; the model can represent three specific setups (see Fig. 1a). 
 
The total reflection amplitude coefficient for any realization of this model is give by 
\begin{equation}
r=-\frac{r_{1}\left(e^{2ikn_{2}l_{2}}r_{2}+1\right)+e^{2ikn_{1}l_{1}}\left(r_{2}+e^{2ikn_{2}l_{2}}\right)}{1+e^{2ikn_{2}l_{2}}r_{2}+e^{2ikn_{1}l_{1}}\left(r_{2}+e^{2ikn_{2}l_{2}}\right)r_{1}}.\label{eq:r}
\end{equation}
To find the EPs of this model analytically 
we impose the two conditions $r(\omega)=r'(\omega)=0$ simultaneously. Defining
$x=e^{2ikn_2l_{2}+2\pi ip},y=e^{2ikn_1l_{1}},$ we get
\begin{align}
x&=\frac{r_{2}^{2}\left(\frac{1}{\Lambda}-1\right)-\frac{1}{\Lambda}-1\pm\sqrt{\left(r_{2}^{2}\left(1-\frac{1}{\Lambda}\right)+1+\frac{1}{\Lambda}\right)^{2}-4r_{2}^{2}}}{2r_{2}},\nonumber\\
y&=r_{1}\frac{r_{2}^{2}\left(\Lambda-1\right)-\Lambda-1\text{\ensuremath{\mp\sqrt{\left(r_{2}^{2}\left(\Lambda-1\right)+\Lambda+1\right)^{2}-4\Lambda^{2}r_{2}^{2}}}}}{2r_{2}},
\end{align}
where $\Lambda\equiv\frac{n_{1}l_{1}}{n_{2}l_{2}}$ is the ratio of the optical lengths of the slabs. By expressing $k=\omega/c$ using $x$ and $y$ and equating, we obtain a general analytic condition for a CPA EP that holds in the weak and strong
coupling regimes, (see SM Sec. V for details):
\begin{equation}
ye^{2\pi ip\Lambda}=x^{\Lambda}.
\end{equation}
 Using this method we can find quite easily continuous curves of EPs in the parameter space of the model (see Fig. S3), allowing design flexibility. Assuming $\Lambda$ is an integer, neglecting dispersion and other non-idealities, the model gives an infinite number of simultaneous EPs for the same parameter values, labeled by an integer $p.$ For $\Lambda=1,$ equal optical length, we obtain analytical solutions for the EPs    
 $$ \frac{\omega}{c}=\frac{\ln\left(\frac{-r_{2}\left(r^\pm_{1}+1\right)}{2}\right)+2p\pi  i}{2in_{1}l_{1}},\,\,r^\pm_{1}=\frac{-r_{2}^{2}+2\pm2\sqrt{1-r_{2}^{2}}}{r_{2}^{2}}.$$
 EPs for virtual CPA with $\Lambda=1$ are shown in Fig 1b. In the generic case, $\Lambda \neq 1$, the solution of Eq. (4) only gives a single EP; such an example is shown in Fig. 1c. Further examples are given in the SM Sec. V.
\begin{figure}
\subfloat[]{\includegraphics[width=7.3cm]{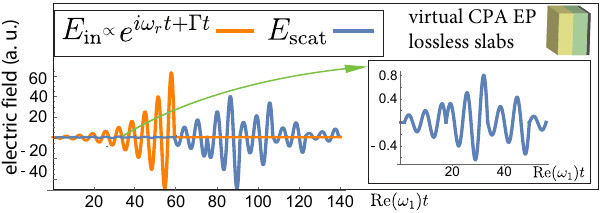}

}

\subfloat[]{\includegraphics[width=7.3cm]{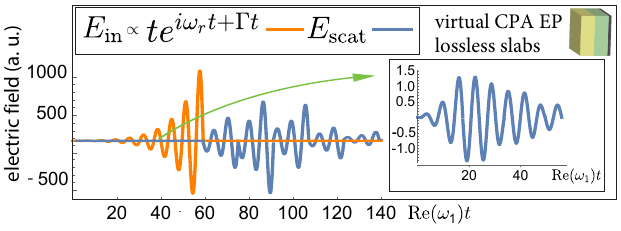}

}

\subfloat[]{\includegraphics[width=7cm]{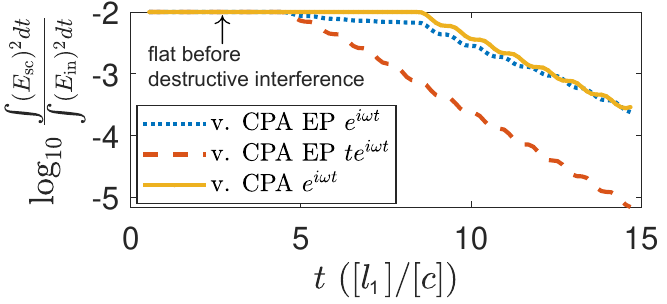}
}
\caption{Scattered field at a virtual CPA EP for the inputs of (a) $e^{i\omega_r t+\Gamma t}$ 
and (b) $te^{i\omega_r t+\Gamma t}$ multiplied by $\left(\theta\left(t-t_{1}\right)-\theta\left(t-t_{2}\right)\right),$
where $\omega_{1}=(1.57+0.575i)/2+\pi$, same parameters as Fig. 1b, except both optical lengths are doubled.
Both inputs captured almost perfectly after an initial transient;
the relative scattering for $e^{i\omega_{r}t}$ is larger, see insets. (c) Relative integrated scattered energies for both inputs compared with a virtual CPA in one cavity with the same total length, Im$(\omega),$ and $r_1.$ \label{fig:Scattered}}
\end{figure}

We now choose an appropriate model for a virtual CPA EP and calculate the temporal response for the relevant finite-time inputs, where $\omega_r=\mathrm{Re}(\omega),\,\,\Gamma=\mathrm{Im}(\omega).$ To that end, we proceed similarly to Ref. \cite{karousos2008transmission}, see SM Sec. VI. We present in Fig. \ref{fig:Scattered} the scattered fields
for the inputs $\exp(i\omega_r t+\Gamma t)$  (a) and $t\exp(i\omega_r t+\Gamma t)$
(b) multiplied by the step functions $\left(\theta\left(t-t_{1}\right)-\theta\left(t-t_{2}\right)\right),$ in a lossless two-slab system similar to that in Fig. 1 (b).
As expected, after a transient, both inputs are not scattered, but 
the relative instantaneous scattering for the input $te^{i\omega_{1}t}$ is much smaller compared to $e^{i\omega_{1}t},$ see insets. In Fig. 2 (c) we present the relative integrated reflected \emph{energies} as functions of time for both inputs in Fig. 2 (a) and (b), compared to $e^{i\omega_{1}t}$ at a virtual CPA in a single-cavity setup with the same total length, Im$(\omega),$ and $r_1.$  Clearly, the impedance matching is superior for both inputs at the virtual CPA EP, and the input $te^{i\omega t}$ performs significantly better. In addition, the total energy accumulated and released in the slab system is larger by two orders of magnitude for $te^{i\omega t}$
(see SM Fig. S5).
\begin{figure}
\subfloat[]{\begin{centering}
\includegraphics[width=6.5cm]{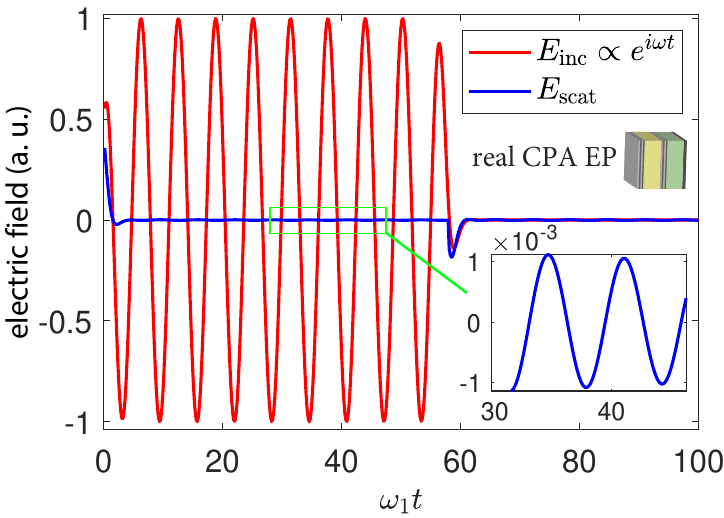}
\par\end{centering}
}\\
\subfloat[]{\begin{centering}
\includegraphics[width=6.5cm]{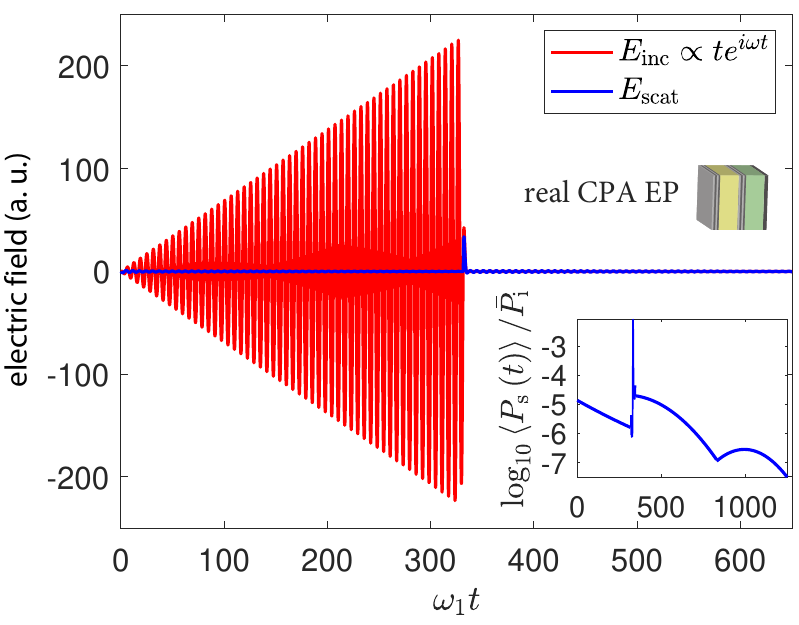}
\par\end{centering}
}
\caption{Scattered wave at a real CPA EP for the inputs of $e^{i\omega_{1}t}$
(a) and $te^{i\omega_{1}t}$ (b) multiplied by $\left(\tanh\left(t\right)+1-\left(\tanh\left(t-t_1\right)+1\right)\right)/2,$
for a high Q cavity with Bragg mirrors and EP parameters in Fig. 1. 
Both inputs are absorbed in steady state. \label{fig:time_domain_two_slabs_Bragg}}
\end{figure}

Now we calculate the temporal responses for CPA EP (real $\omega$) in a two-slab setup with loss. Here we seek efficient transduction, so we must keep track of energy which is scattered after the input is turned off. To alleviate several computational challenges, we developed a general approach to perform a numerical inverse FT of the output, see SM Sec. VI B. Since the FT of a step function decays slowly at high frequencies, we used $\tanh(t)$ as a smooth
switching function. Importantly, we show there that a term in the FT of $\tilde{\theta}(t-t_1)t\exp(i\omega t)$ is proportional to the switching-off time $t_1$. This implies that for $t\exp(i\omega t)$ there is negligible accumulation of unabsorbed energy at a CPA EP (due to the double zero), whereas for 
the conversion process, at an ordinary CPA there would be a large accumulation of unabsorbed energy in the resonator which will be emitted after turn off.

In Fig. \ref{fig:time_domain_two_slabs_Bragg} we present the temporal responses to $\exp(i\omega t)$ and $t\exp(i\omega t)$ at a real CPA EP 
in a high-Q two-slab setup with Bragg mirrors. In a high-Q cavity since the reflection is large and the absorption is low, the scattered field and equilibration time are larger.
Evidently, at the CPA EP, the scattering from both input signals is relatively small and does decay in time while the pulse is on, as seen in the inset of Fig. 3 (b) (the moving-average of the scattered power normalized by the average incident power), implying large  dissipation within the medium. For the increasing input $t\exp(i\omega t)$ it is noteworthy that negligible scattering occurs after turn off of the input, in agreement with our conclusion above. 
This contrasts with the response at ordinary CPA (single zero) shown in Fig. 4 in a low-Q two-slab system without Bragg mirrors. Here, $\exp(i\omega t)$ is absorbed as expected and $te^{i\omega_{1}t}$ is converted to $e^{i\omega_{1}t}$ in agreement with our CW analysis.
Moreover, the scattered field in response to $e^{i\omega_{1}t}$ at the CPA is larger than at the CPA EP even though the Q-factor of the CPA EP setup is much larger, which means that the effective transient scattering at a CPA EP is much smaller (approximately 2 orders of magnitude for the times in the insets in Figs. 3 (a) and 4 (a)). Finally, after stopping the signal for the input $te^{i\omega_{1}t},$ the substantial accumulated energy is released, in agreement with our analysis above, see Fig. 4 (b).
\begin{figure}
\subfloat[]{\begin{centering}
\includegraphics[width=6.5cm]{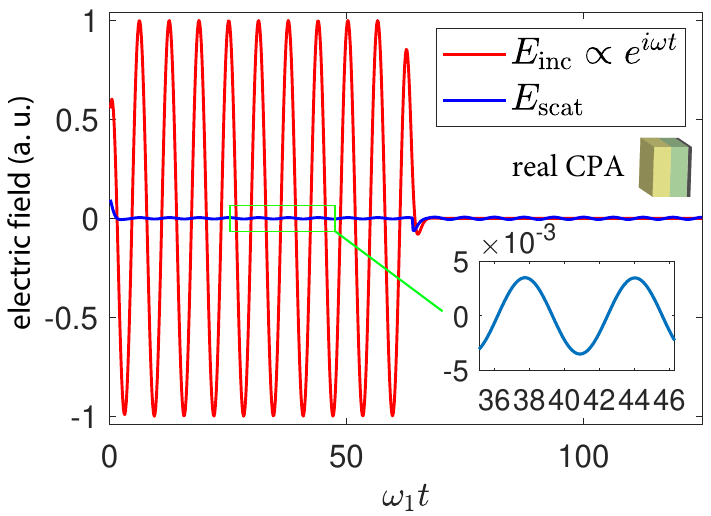}
\par\end{centering}
}

\subfloat[]{\begin{centering}
\includegraphics[width=6.5cm]{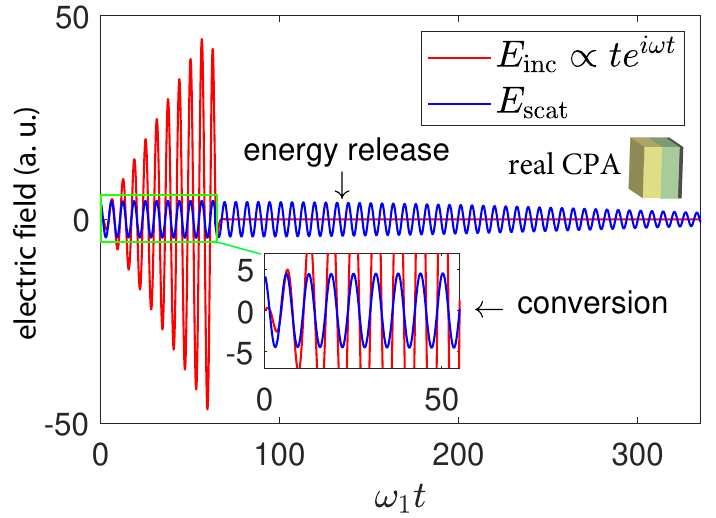}
\par\end{centering}
}

\caption{Scattered wave at a real CPA (not an EP) in a low-Q two-slab setup for the inputs  $e^{i\omega_{1}t}$ (a)
and $te^{i\omega_{1}t}$ (b) multiplied by $\left(\tanh\left(t\right)+1-\left(\tanh\left(t-50\right)+1\right)\right)/2,$
with $l_{1}=1,l_{2}=1.5,n_{1}=1.533,n_{2}=1.4+0.3i,$ $\omega_{\mathrm{CPA}}=1.27.$ It can be seen that
$te^{i\omega_{1}t}$ is converted to $e^{i\omega_{1}t}$ (constant amplitude)
and there is emission of the accumulated energy after the input stops. \label{fig:temporal_domain_CPA}}
\end{figure}

\begin{figure}
\includegraphics[width=0.5\textwidth]{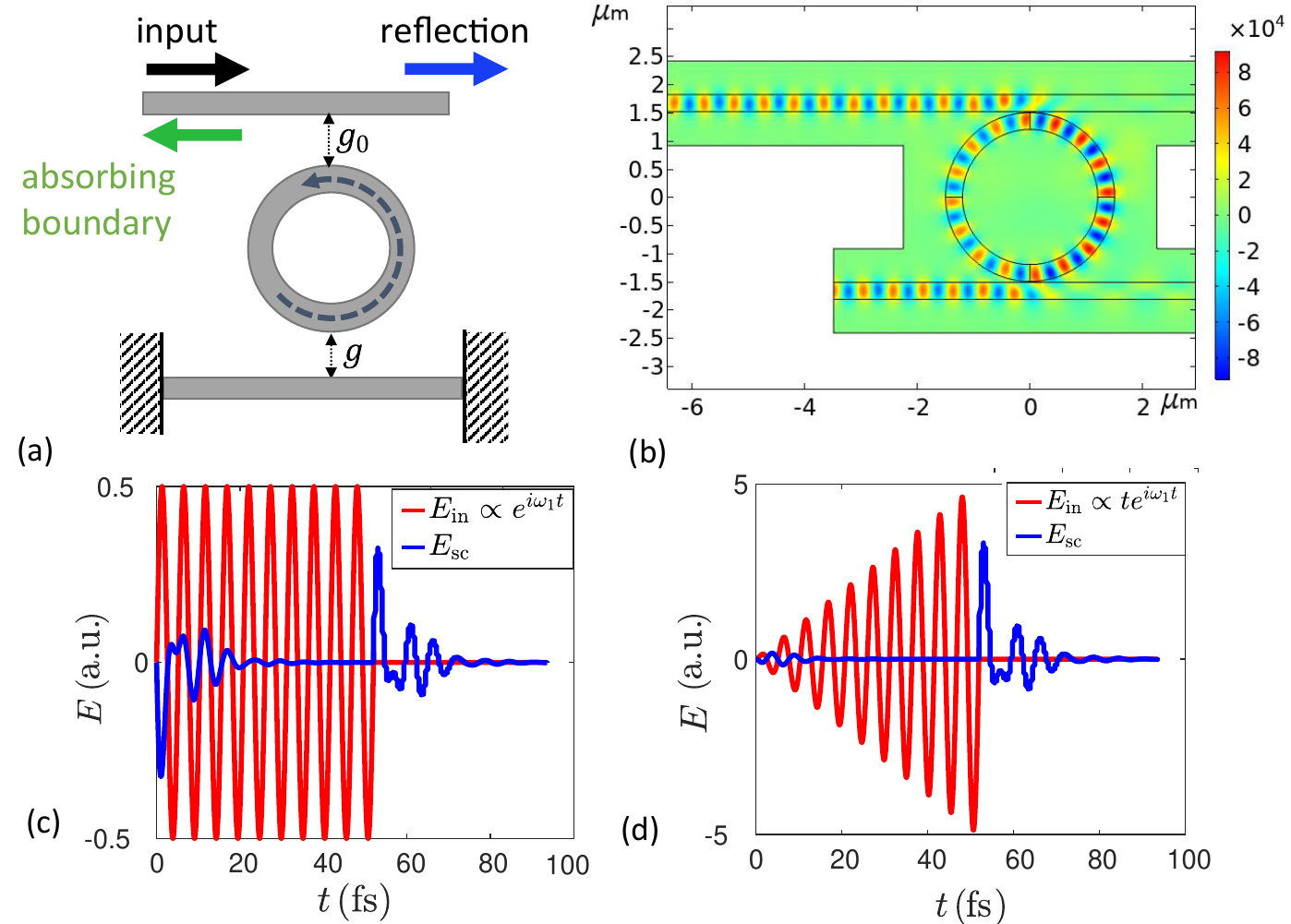}
\caption{(a) Photonic integrated circuit (PIC) platform to realize CPA EP on real axis.  (b) Electric field in the PIC at the CPA EP. (c),(d) temporal response at the EP for constant and linearly growing input pulses. \label{fig:example_alu}}
\end{figure}

We validate our model analysis by simulating a realistic setup of a photonic integrated circuit (PIC) that can be easily tuned to a CPA EP on the real axis, and verify its unique scattering features with full-wave simulations. The PIC consists of coupled ring-optical waveguide (CROW) resonators with a geometry that can be readily implemented using conventional PIC technology, see Fig. \ref{fig:example_alu}(a). The coupling parameters between the waveguides and the ring resonator $g$ and $g_0,$ which are determined by their distances, allow us to tune the setup to a CPA EP, and in this platform they can be tuned with high accuracy using a nanopositioner stage, making it highly promising for an experimental implementation. The parameters of the setup were chosen to have the CPA EP close to the optical telecommunication wavelength $ \lambda_0 = 1.56\mu\mathrm{m},$ see SM, Sec. VIII. In Fig. 5 (b) we plot the electric field distribution for a sinusoidal excitation and input power $P_{\mathrm{in}} = 1$(W/m) at a CPA EP occurring at $g/g_0=1.08.$ As expected, there is no reflection at the right port. Finally, we plot the time domain response for both inputs in Fig. 5 (c) and (d), showcasing the same dynamics predicted in the three-mirror model.

We conclude that cavities tuned to an EP provide superior performance for wave capture and impedance matching, for both energy storage (virtual case) and transduction (coherent absorption). Particularly interesting is the possibility of perfectly absorbing a linearly growing wave as well as a CW wave, which hasn't been previously explored. The generalization of this work to higher-degeneracy EPs was mentioned above and is straightforward. While we have only shown single port realizations of absorbing structures tuned to an EP, our conclusions are valid for multiport excitation, as long as the appropriate wavefront is imposed at each port.

\section{Acknowledgements}
We acknowledge the fruitful discussions with Tsampikos Kottos. This work was supported by a grant from the Simons Foundation. AA and AM acknowledge a grant from the Air Force Office of Scientific Research.
\FloatBarrier
\bibliographystyle{unsrt}

\begin{thebibliography}{10}

\bibitem{chong2010coherent}
YD~Chong, Li~Ge, Hui Cao, and A~Douglas Stone.
\newblock Coherent perfect absorbers: time-reversed lasers.
\newblock {\em Physical review letters}, 105(5):053901, 2010.

\bibitem{baranov2017coherent}
Denis~G Baranov, Alex Krasnok, and Andrea Alu.
\newblock Coherent virtual absorption based on complex zero excitation for
  ideal light capturing.
\newblock {\em Optica}, 4(12):1457--1461, 2017.

\bibitem{el2007theory}
R~El-Ganainy, KG~Makris, DN~Christodoulides, and Ziad~H Musslimani.
\newblock Theory of coupled optical pt-symmetric structures.
\newblock {\em Optics letters}, 32(17):2632--2634, 2007.

\bibitem{makris2008beam}
Konstantinos~G Makris, R~El-Ganainy, DN~Christodoulides, and Ziad~H Musslimani.
\newblock Beam dynamics in pt-symmetric optical lattices.
\newblock {\em Physical Review Letters}, 100(10):103904, 2008.

\bibitem{wiersig2011structure}
Jan Wiersig.
\newblock Structure of whispering-gallery modes in optical microdisks perturbed
  by nanoparticles.
\newblock {\em Physical Review A}, 84(6):063828, 2011.

\bibitem{wiersig2014chiral}
Jan Wiersig.
\newblock Chiral and nonorthogonal eigenstate pairs in open quantum systems
  with weak backscattering between counterpropagating traveling waves.
\newblock {\em Physical Review A}, 89(1):012119, 2014.

\bibitem{kato2013perturbation}
Tosio Kato.
\newblock {\em Perturbation theory for linear operators}, volume 132.
\newblock Springer Science \& Business Media, 2013.

\bibitem{moiseyev2011non}
Nimrod Moiseyev.
\newblock {\em Non-Hermitian quantum mechanics}.
\newblock Cambridge University Press, 2011.

\bibitem{bender1998real}
Carl~M Bender and Stefan Boettcher.
\newblock Real spectra in non-hermitian hamiltonians having p t symmetry.
\newblock {\em Physical Review Letters}, 80(24):5243, 1998.

\bibitem{guo2009observation}
A~Guo, GJ~Salamo, D~Duchesne, R~Morandotti, M~Volatier-Ravat, V~Aimez,
  GA~Siviloglou, and DN~Christodoulides.
\newblock Observation of p t-symmetry breaking in complex optical potentials.
\newblock {\em Physical review letters}, 103(9):093902, 2009.

\bibitem{ruter2010observation}
Christian~E R{\"u}ter, Konstantinos~G Makris, Ramy El-Ganainy, Demetrios~N
  Christodoulides, Mordechai Segev, and Detlef Kip.
\newblock Observation of parity--time symmetry in optics.
\newblock {\em Nature physics}, 6(3):192--195, 2010.

\bibitem{heiss2012physics}
WD~Heiss.
\newblock The physics of exceptional points.
\newblock {\em Journal of Physics A: Mathematical and Theoretical},
  45(44):444016, 2012.

\bibitem{doppler2016dynamically}
J{\"o}rg Doppler, Alexei~A Mailybaev, Julian B{\"o}hm, Ulrich Kuhl, Adrian
  Girschik, Florian Libisch, Thomas~J Milburn, Peter Rabl, Nimrod Moiseyev, and
  Stefan Rotter.
\newblock Dynamically encircling an exceptional point for asymmetric mode
  switching.
\newblock {\em Nature}, 537(7618):76--79, 2016.

\bibitem{xu2016topological}
Haitan Xu, David Mason, Luyao Jiang, and JGE Harris.
\newblock Topological energy transfer in an optomechanical system with
  exceptional points.
\newblock {\em Nature}, 537(7618):80--83, 2016.

\bibitem{feng2017non}
Liang Feng, Ramy El-Ganainy, and Li~Ge.
\newblock Non-hermitian photonics based on parity--time symmetry.
\newblock {\em Nature Photonics}, 11(12):752--762, 2017.

\bibitem{zhang2018hybrid}
Xu-Lin Zhang and Che~Ting Chan.
\newblock Hybrid exceptional point and its dynamical encircling in a two-state
  system.
\newblock {\em Physical Review A}, 98(3):033810, 2018.

\bibitem{pick2017general}
Adi Pick, Bo~Zhen, Owen~D Miller, Chia~W Hsu, Felipe Hernandez, Alejandro~W
  Rodriguez, Marin Solja{\v{c}}i{\'c}, and Steven~G Johnson.
\newblock General theory of spontaneous emission near exceptional points.
\newblock {\em Optics express}, 25(11):12325--12348, 2017.

\bibitem{zhang2019quantum}
Mengzhen Zhang, William Sweeney, Chia~Wei Hsu, Lan Yang, AD~Stone, and Liang
  Jiang.
\newblock Quantum noise theory of exceptional point amplifying sensors.
\newblock {\em Physical review letters}, 123(18):180501, 2019.

\bibitem{zhen2015spawning}
Bo~Zhen, Chia~Wei Hsu, Yuichi Igarashi, Ling Lu, Ido Kaminer, Adi Pick,
  Song-Liang Chua, John~D Joannopoulos, and Marin Solja{\v{c}}i{\'c}.
\newblock Spawning rings of exceptional points out of dirac cones.
\newblock {\em Nature}, 525(7569):354--358, 2015.

\bibitem{chen2020revealing}
Hua-Zhou Chen, Tuo Liu, Hong-Yi Luan, Rong-Juan Liu, Xing-Yuan Wang, Xue-Feng
  Zhu, Yuan-Bo Li, Zhong-Ming Gu, Shan-Jun Liang, He~Gao, et~al.
\newblock Revealing the missing dimension at an exceptional point.
\newblock {\em Nature Physics}, 16(5):571--578, 2020.

\bibitem{wang2020electromagnetically}
Changqing Wang, Xuefeng Jiang, Guangming Zhao, Mengzhen Zhang, Chia~Wei Hsu,
  Bo~Peng, A~Douglas Stone, Liang Jiang, and Lan Yang.
\newblock Electromagnetically induced transparency at a chiral exceptional
  point.
\newblock {\em Nature Physics}, 16(3):334--340, 2020.

\bibitem{naghiloo2019quantum}
M~Naghiloo, M~Abbasi, Yogesh~N Joglekar, and KW~Murch.
\newblock Quantum state tomography across the exceptional point in a single
  dissipative qubit.
\newblock {\em Nature Physics}, 15(12):1232--1236, 2019.

\bibitem{song2021plasmonic}
Qinghua Song, Mutasem Odeh, Jes{\'u}s Z{\'u}{\~n}iga-P{\'e}rez, Boubacar
  Kant{\'e}, and Patrice Genevet.
\newblock Plasmonic topological metasurface by encircling an exceptional point.
\newblock {\em Science}, 373(6559):1133--1137, 2021.

\bibitem{miri2019exceptional}
Mohammad-Ali Miri and Andrea Alu.
\newblock Exceptional points in optics and photonics.
\newblock {\em Science}, 363(6422):eaar7709, 2019.

\bibitem{sweeney2019perfectly}
William~R Sweeney, Chia~Wei Hsu, Stefan Rotter, and A~Douglas Stone.
\newblock Perfectly absorbing exceptional points and chiral absorbers.
\newblock {\em Physical review letters}, 122(9):093901, 2019.

\bibitem{wang2021coherent}
Changqing Wang, William~R Sweeney, A~Douglas Stone, and Lan Yang.
\newblock Coherent perfect absorption at an exceptional point.
\newblock {\em Science}, 373(6560):1261--1265, 2021.

\bibitem{heugel2010analogy}
Simon Heugel, Alessandro~S Villar, Markus Sondermann, Ulf Peschel, and Gerd
  Leuchs.
\newblock On the analogy between a single atom and an optical resonator.
\newblock {\em Laser Physics}, 20(1):100--106, 2010.

\bibitem{wenner2014catching}
J~Wenner, Yi~Yin, Yu~Chen, R~Barends, B~Chiaro, E~Jeffrey, J~Kelly, A~Megrant,
  JY~Mutus, C~Neill, et~al.
\newblock Catching time-reversed microwave coherent state photons with 99.4\%
  absorption efficiency.
\newblock {\em Physical Review Letters}, 112(21):210501, 2014.

\bibitem{sweeney2020theory}
William~R Sweeney, Chia~Wei Hsu, and A~Douglas Stone.
\newblock Theory of reflectionless scattering modes.
\newblock {\em Physical Review A}, 102(6):063511, 2020.

\bibitem{dhia2018trapped}
Anne-Sophie Bonnet-Ben Dhia, Lucas Chesnel, and Vincent Pagneux.
\newblock Trapped modes and reflectionless modes as eigenfunctions of the same
  spectral problem.
\newblock {\em Proceedings of the Royal Society A: Mathematical, Physical and
  Engineering Sciences}, 474(2213):20180050, 2018.

\bibitem{chen2017exceptional}
Weijian Chen, {\c{S}}ahin Kaya~{\"O}zdemir, Guangming Zhao, Jan Wiersig, and
  Lan Yang.
\newblock Exceptional points enhance sensing in an optical microcavity.
\newblock {\em Nature}, 548(7666):192--196, 2017.

\bibitem{park2020symmetry}
Jun-Hee Park, Abdoulaye Ndao, Wei Cai, Liyi Hsu, Ashok Kodigala, Thomas
  Lepetit, Yu-Hwa Lo, and Boubacar Kant{\'e}.
\newblock Symmetry-breaking-induced plasmonic exceptional points and nanoscale
  sensing.
\newblock {\em Nature Physics}, 16(4):462--468, 2020.

\bibitem{lai2019observation}
Yu-Hung Lai, Yu-Kun Lu, Myoung-Gyun Suh, Zhiquan Yuan, and Kerry Vahala.
\newblock Observation of the exceptional-point-enhanced sagnac effect.
\newblock {\em Nature}, 576(7785):65--69, 2019.

\bibitem{wu2019observation}
Yang Wu, Wenquan Liu, Jianpei Geng, Xingrui Song, Xiangyu Ye, Chang-Kui Duan,
  Xing Rong, and Jiangfeng Du.
\newblock Observation of parity-time symmetry breaking in a single-spin system.
\newblock {\em Science}, 364(6443):878--880, 2019.

\bibitem{karousos2008transmission}
Anastasios Karousos, George Koutitas, and Costas Tzaras.
\newblock Transmission and reflection coefficients in time-domain for a
  dielectric slab for uwb signals.
\newblock In {\em VTC Spring 2008-IEEE Vehicular Technology Conference}, pages
  455--458. IEEE, 2008.

\end{thebibliography}

\end{document}


\title{Supplementary Material: The signature of absorbing exceptional points in the time domain}
\author{Asaf Farhi\textsuperscript{1} , Ahmed Mekawy\textsuperscript{3,4} , Andrea Alu\textsuperscript{3,4,5}, and Douglas Stone\textsuperscript{1,2}}
\affiliation{\textsuperscript{1}Department of Applied Physics, Yale University, New Haven, Connecticut
06520, USA}\vspace*{2cm}
\affiliation{\textsuperscript{2}Yale Quantum Institute, Yale University, New Haven, Connecticut 06520,
USA}
\affiliation{\textsuperscript{3} Photonics Initiative, Advanced Science Research Center, City University
of New York, New York, New York 10031, USA}
\affiliation{\textsuperscript{4}Department of Electrical Engineering, City College of The City University
of New York, New York, New York 10031, }
\affiliation{\textsuperscript{5}Physics Program, Graduate Center, City University of New York, New
York, New York 10016, USA}

\maketitle

\renewcommand{\thefigure}{S\arabic{figure}}

\section{Conversions and response to the input $t^{2}e^{i\omega_{n}t}$ }

We examine the temporal response to the input $t^{2}e^{i\omega_{n}t}$
\begin{gather}
\mathcal{F}\left(t^{2}e^{i\omega_{n}t}\right)=\delta^{\left(2\right)}\left(\omega-\omega_{n}\right),\nonumber\\
\int\delta^{\left(2\right)}(\omega-\omega_{n})\lambda\left(\omega\right)e^{i\omega t}d\omega=\left(\lambda'\left(\omega_{n}\right)e^{i\omega_{n}t}+\lambda\left(\omega_{n}\right)ite^{i\omega_{n}t}\right)^{'}=\nonumber\\
\lambda^{\left(2\right)}\left(\omega_{n}\right)e^{i\omega_{n}t}+\lambda'\left(\omega_{n}\right)ite^{i\omega_{n}t}+\lambda'\left(\omega_{n}\right)ite^{i\omega_{n}t}+\lambda\left(\omega_{n}\right)ie^{i\omega_{n}t}-\lambda\left(\omega_{n}\right)t^{2}e^{i\omega_{n}t},\nonumber
\end{gather}
which vanishes at a second order CPA EP where $\lambda\left(\omega_{n}\right)=\lambda'\left(\omega_{n}\right)=\lambda^{\left(2\right)}\left(\omega_{n}\right)=0.$
This leads us to the possibility of converting between $e^{i\omega_{n}t}t^{m}$
signals. For example, for a standard CPA and the input $te^{i\omega_{n}t}$
we get an output of the form $e^{i\omega_{n}t}$
\[
\int\delta'(\omega-\omega_{n})\lambda\left(\omega\right)e^{i\omega t}d\omega=\lambda'\left(\omega_{n}\right)e^{i\omega_{n}t}.
\]
 Similarly, a CPA EP converts $e^{i\omega_{n}t}t^{2}\rightarrow e^{i\omega_{n}t}$
since $\lambda^{'}\left(\omega_{n}\right)=\lambda\left(\omega_{n}\right)=0,\lambda^{\left(2\right)}\left(\omega_{n}\right)\neq0.$
When $\lambda^{\left(2\right)}\left(\omega_{n}\right)=\lambda\left(\omega_{n}\right)=0,\lambda'\left(\omega_{n}\right)\neq0,$
we get the conversion $e^{i\omega_{n}t}t^{2}\rightarrow te^{i\omega_{n}t},$ see Fig. S1.
\section{Solution to the wave equation}
For concreteness we consider the electromagnetic wave equation. To
satisfy the wave equation $\frac{\partial^{2}E}{\partial z^{2}}-\frac{n_{i}^{2}}{c^{2}}\frac{\partial^{2}E}{\partial t^{2}}=0\text{,}$ where
$n_{i}$ is the refractive index, we look for an electric field $E$
that is a function of $\frac{ct}{n_{i}}-z.$ In order for $E$ to
be a function of the form $t^{m}e^{i\omega t}$ at a fixed location
we write
\[
E=\left(\frac{c}{n_{i}}t-z\right)^{m}e^{i\left(-\omega t+kn_{i}z\right)}\theta\left(\frac{c}{n_{i}}t-z\right)\,\,\,t\ge0,
\]
where $\theta$ is a step function. This also ensures that closer to the
surface current that generates the field $\boldsymbol{J}\propto t^{m}e^{i\omega_{n}t}\delta\left(z-z_{0}\right)\boldsymbol{e}_{y}$
we have a larger field amplitude since the field was generated at
a later time.

In order to satisfy the wave equation
for a complex $\omega, k\equiv k_{r}+ik_{i}$ has to be complex, and we obtain the first wave-equation solution
$$
E=e^{-i\omega_{r}t+\Gamma t+ik_{r}z-k_{i}z},
$$
which decays exponentially in space, and the second wave-equation solution
$$E=k\left(vt-z\right)e^{-i\omega_{r}t+\Gamma t+ik_{r}z-k_{i}z}.$$
\begin{figure}
\begin{centering}
\includegraphics[scale=0.35]{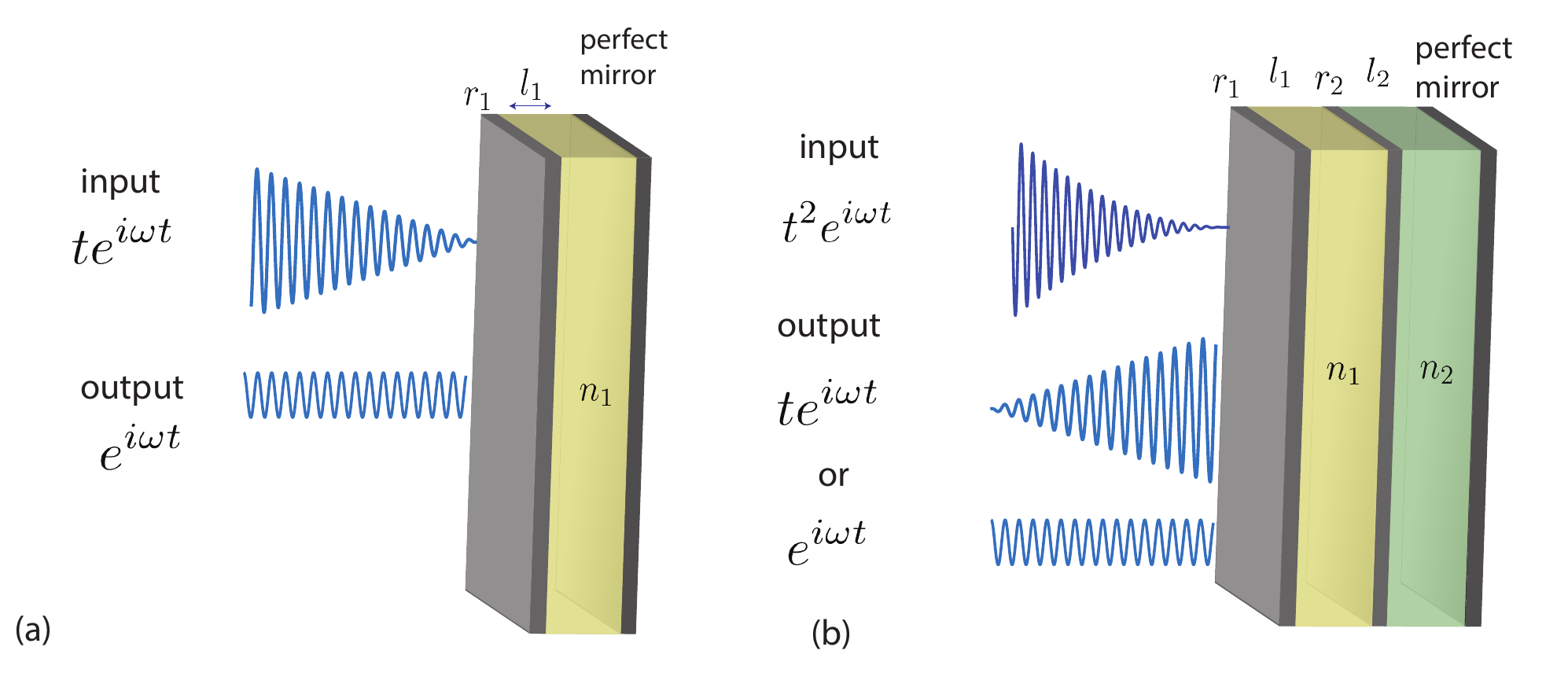}
\par\end{centering}
\caption{ Conversion between $te^{i\omega t}$
and $e^{i\omega t}$ at a real CPA in a slab setup (a) and between
$t^{2}e^{i\omega t}$ and $te^{i\omega t}$ or $e^{i\omega t}$ in
a two slab setup (b).\label{fig:The-setup-of}}

\end{figure}
\section{Calculating the reflection coefficient}
We write the CW field with incoming boundary conditions
\begin{equation}
\left\{ \begin{array}{cc}
A_{1}e^{ikx}+A_{2}e^{-ikx} & x\in\mathrm{I}\\
B_{1}e^{ikn_{1}x}+B_{2}e^{-ikn_{1}x} & x\in\mathrm{II}\\
C_{1}e^{ikn_{2}x}+C_{2}e^{-ikn_{2}x} & x\in\mathrm{III}
\end{array}\right..
\end{equation}
We then impose boundary conditions
\begin{gather}
\frac{r_{1}}{t_{1}}A_{1}e^{ik\left(-l_{1}\right)}+\frac{1}{t_{1}}A_{2}e^{-ik\left(-l_{1}\right)}=B_{2}e^{-ikn_{1}\left(-l_{1}\right)},\nonumber\\
\frac{1}{t_{1}}A_{1}e^{ik\left(-l_{1}\right)}+\frac{r_{1}}{t_{1}}A_{2}e^{-ik\left(-l_{1}\right)}=B_{1}e^{ikn_{1}\left(-l_{1}\right)},\nonumber\\
\frac{r_{2}}{t_{2}}B_{1}+\frac{1}{t_{2}}B_{2}=C_{2},\nonumber\\
\frac{1}{t_{2}}B_{1}+\frac{r_{2}}{t_{2}}B_{2}=C_{1},\nonumber\\
-C_{1}e^{2ikn_{2}l_{2}}=C_{2},\nonumber
\end{gather}
where $r_{1}=\frac{n_{1}-\left(\frac{n_{4}}{n_{3}}\right)^{2N_{1}}}{n_{1}+\left(\frac{n_{4}}{n_{3}}\right)^{2N_{1}}},\,\,\,r_{2}=\frac{n_{2}-\left(\frac{n_{6}}{n_{5}}\right)^{2N_{2}}n_{1}}{n_{2}+\left(\frac{n_{6}}{n_{5}}\right)^{2N_{2}}n_{1}},$
$n_{3},n_{4}$ and $n_{5},n_{6}$ are the refractive indices in the
left and right Bragg mirrors, and $N_{1}$ and $N_{2}$ are the number
of repetitions in the left and right Bragg mirrors respectively ($N_{1}=N_{2}=0$
for a two slab system without Bragg mirrors).

\section{Calculating virtual CPA EPs}

We consider lossless slabs for the special case of equal optical lengths
of the slabs, which admits an analytical solution. We assume general
reflection coefficients $r_{1},r_{2}$ and for simplicity also vacuum
in both slabs $n_{1}=n_{2}=1$ and therefore $l_{1}=l_{2}.$ We will
then map this setup to a setup of two slabs that results in the reflection
coefficients that we will calculate. 

In order to calculate the EP analytically we first substitute $x=e^{2ikl_{1}+2\pi pi}=e^{2ikl_{2}+2\pi qi}$
and get a polynomial for the numerator of $r.$ We impose no reflection
and get
$$
\left(e^{2ikl_{1}}-x_{1}\right)\left(e^{2ikl_{2}}-x_{2}\right)=r_{2}x+r_{1}+x^{2}+r_{1}r_{2}x=0,\nonumber
$$
with the solutions
\[
x_{1,2}=\frac{-r_{1}r_{2}-r_{2}\pm\sqrt{\left(r_{1}r_{2}+r_{2}\right)^{2}-4r_{1}}}{2},\,\,\,k=\frac{\ln x-2\pi pi}{2in_{1}l_{1}}.
\]
We find the EP parameters by imposing vanishing of $\Delta$ of the
CPA polynomial
\begin{gather}
\left[r_{2}\left(1+r_{1}\right)\right]^{2}=4r_{1},\nonumber\\
r_{2}^{2}+2r_{2}^{2}r_{1}-4r_{1}+r_{2}^{2}r_{1}^{2}=0,\nonumber
\end{gather}
and obtain the EP relation
\[
r_{1_{1,2}}=\frac{-r_{2}^{2}+2\pm2\sqrt{1-r_{2}^{2}}}{r_{2}^{2}},
\]
which is satisfied for an infinite number of EPs occurring simultaneously
due to the multiplicity in $p.$ Interestingly, when increasing $l_{1},$
$\Delta\omega=\frac{\pi c}{n_{1}l_{1}}$ decreases, which may result
in negligible dispersions in $n_{1}\left(\omega\right),n_{2}\left(\omega\right)$
at adjacent frequencies and possibly realizable experimental implementation.
Also, when considering real CPA EPs, the high-Q or large $l_{1}$
limits may imply $\mathrm{Im}\left(n_{1}\right),\mathrm{Im}\left(n_{2}\right)\ll1$
and $\Delta k$ that is approximately real, and thus real $x$ or
$p,l_{1}\gg1,$ may allow this phenomenon to approximately occur.
Importantly, in the case of an RSM EP $\mathrm{Im}\left(n_{1}\right)=\mathrm{Im}\left(n_{2}\right)=0$
can occur and thus this phenomenon may be realized.

We choose 
\[
r_{1\,\mathrm{EP}}=0.1,\,\,\,r_{2\mathrm{\,EP}}=0.57496.
\]
In Fig. 1 (b) we calculate $ks$ of the virtual
CPA from the general CPA equation close to the CPA EP. It can be seen
that since the solution is analytical we get an exact EP.

\begin{figure}
\subfloat[]{
\includegraphics[width=8cm]{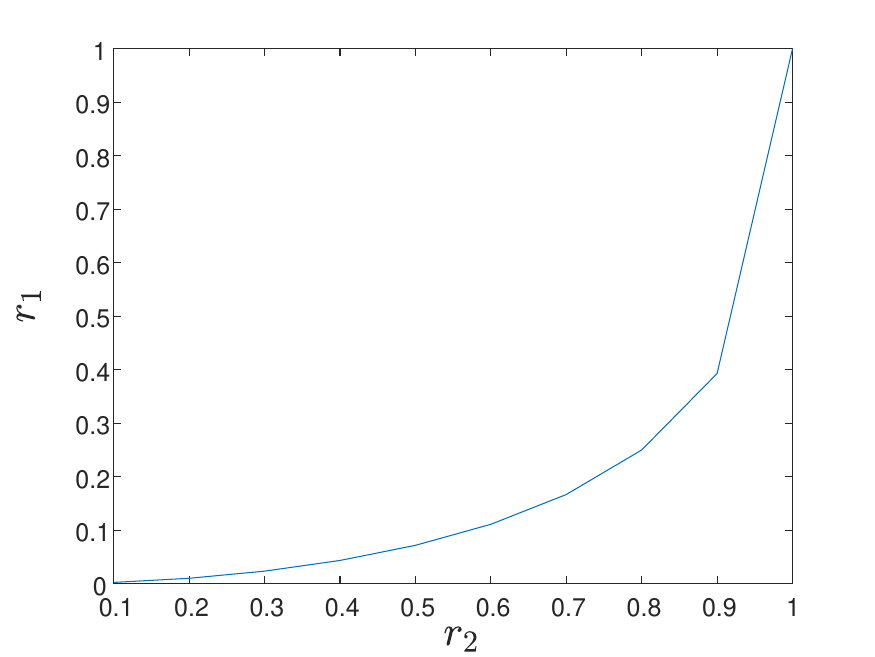}}
\subfloat[]{
\includegraphics[width=8cm]{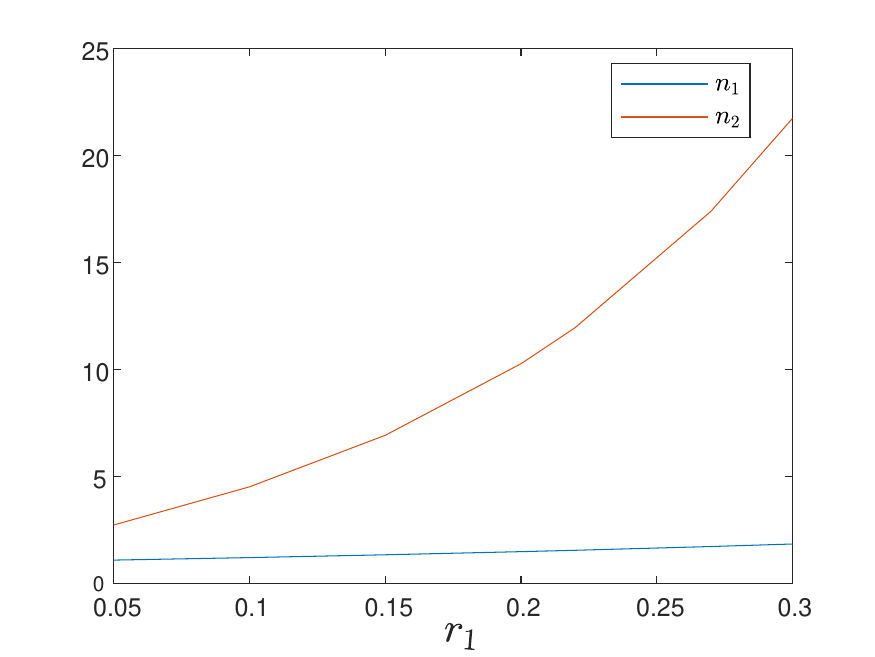}}
\caption{(a) $r_{1}$ as a function of $r_{2}$ for a virtual CPA EP. (b) $n_{1},n_{2}$ as functions of $r_{1}$ for a virtual CPA EP \label{fig:r1r2n1n2}}
\end{figure}

Now we map the mirror setup to a two slab setup. We relate the reflection
coefficient to the refractive indices of the slabs (see Fig. 1 (a)) using
the Fresnel equation relation, which we verify from the first two
boundary conditions
\[
-r_{1/2}=\frac{n_{i}-n_{t}}{n_{i}+n_{t}}.
\]
We then substitute $n=1$ in the host medium and get
\[
n_{1\,\,\mathrm{EP}}=1.22222,n_{2\,\,\mathrm{EP}}=4.52878.
\]
We also calculate $r_{1\,\mathrm{EP}}$ as we vary $r_{2}$ and translate
the results to the slab setup. In Fig.  \ref{fig:r1r2n1n2} (a) we plot $r_{1}\left(r_{2}\right)$ for virtual CPA EP with $n_{1}=n_{2}=1$ and
 in Fig. \ref{fig:r1r2n1n2} (b) we present the mapping of the relation to $n_{1}$ and $n_{2}.$ 

\subsection*{Calculating a virtual CPA }

Here we calculate a virtual CPA in a single-slab setup with the same $\mathrm{Im}\left(\omega\right),r_{1}$
and total length as in Fig. 2.

We first write the continuous-wave (CW) solution in the a one-slab
setup terminated by a perfect mirror

\[
\left\{ \begin{array}{cc}
A_{1}e^{ikx}+A_{2}e^{-ikx} & x\in\mathrm{I}\\
B_{1}e^{ikn_{1}x}+B_{2}e^{-ikn_{1}x} & x\in\mathrm{II}
\end{array}\right..
\]
We impose boundary conditions
\[
\frac{r_{1}}{t_{1}}A_{1}+\frac{1}{t_{1}}A_{2}=B_{2},
\]
\[
\frac{1}{t_{1}}A_{1}+\frac{r_{1}}{t_{1}}A_{2}=B_{1},
\]
\[
-B_{1}e^{2ikn_{1}l_{1}}=B_{2},
\]
 get
\[
\frac{r_{1}}{t_{1}}A_{1}+\frac{1}{t_{1}}A_{2}=-\left(\frac{1}{t_{1}}A_{1}+\frac{r_{1}}{t_{1}}A_{2}\right)e^{2ikn_{1}l_{1}},
\]
\[
r_{1}A_{1}+A_{2}=-\left(A_{1}+r_{1}A_{2}\right)e^{2ikn_{1}l_{1}},
\]
\[
A_{1}\left(r_{1}+e^{2ikn_{1}l_{2}}\right)=-A_{2}\left(1+r_{1}e^{2ikn_{1}l_{1}}\right),
\]
and obtain the reflection coefficient
\[
r=-\frac{r_{1}+e^{2ikn_{1}l_{2}}}{1+r_{1}e^{2ikn_{1}l_{1}}}.
\]
We impose $r=0$
\[
-r_{1}=e^{2ikn_{1}l_{1}+2\pi pi},
\]
and express $k_{\mathrm{CPA}}$
\[
\log\left(-r_{1}\right)=2ikn_{1}l_{1}+2\pi pi,
\]
\[
k_{\mathrm{CPA}}=\frac{\log\left(-r_{1}\right)+2\pi pi}{2in_{1}l_{1}}.
\]

For a single cavity with $n_{1}l_{1}=2$ from the correspondence with
the expression for $k_{\mathrm{CPA\,EP}}$ 
\[
r_{1\,\mathrm{CPA}}\rightarrow r_{2\,\mathrm{EP}}\frac{\left(1+r_{1\,\mathrm{EP}}\right)}{2},\,r_{1\,\mathrm{EP}}=0.1,\,\,\,r_{2\,\mathrm{EP}}=0.57496
\]
we require the same $k$ and get
\[
r_{1\,\mathrm{CPA}}=0.3162,
\]
\[
k_{\mathrm{CPA}}=k_{\mathrm{CPA\,EP}}=\frac{-1.17+3.14i+4\pi i}{4i}=\left(1.57+0.57i\right)/2+\pi.
\]
Alternatively, for $r_{1}=0.1$ and length $n_{2}l_{2}=4$ we get
\[
k_{\mathrm{CPA}}=\frac{\log\left(-0.1\right)+2\pi pi}{8i}=3.53429+0.287823i,
\]
which has the same $\mathrm{Im}(\omega),\,r_1,$ and total length.

\section{Calculating real CPA EPs}
We consider an $m-\mathrm{order}$ EP and define in the transfer function
$\rho\left(\omega\right)=r=f\left(\omega\right)\left(\omega-\omega_{n}\right)^{m+1},$
$\left(\omega-\omega_{n}\right)^{m+1}$ as the numerator and $f\left(\omega\right)=1/\mathrm{denominator.}$
We then conclude that $\frac{\partial^{j}\left(\omega-\omega_{n}\right)^{m+1}}{\partial\omega^{j}}=0\,\,\,\forall j\leq m\rightarrow\frac{\partial^{j}\rho\left(\omega\right)}{\partial\omega^{j}}=0\,\,\,\forall j\leq m.$
We consider a first order EP and impose vanishing of the numerator
of $r$ and its derivative
\begin{gather}
r_{\mathrm{num}}=r_{1}\left(e^{2ikn_{2}l_{2}}r_{2}+1\right)+e^{2ikn_{1}l_{1}}\left(r_{2}+e^{2ikn_{2}l_{2}}\right)=0,\label{eq:r_num}\\
\frac{\partial r_{\mathrm{num}}}{\partial k}=r_{1}n_{2}l_{2}e^{2ikn_{2}l_{2}}r_{2}+e^{2ikn_{1}l_{1}}\left[r_{2}n_{1}l_{1}+\left(n_{1}l_{1}+n_{2}l_{2}\right)e^{2ikn_{2}l_{2}}\right]=0.
\end{gather}
We define

\begin{equation}
x=e^{2ikn_{2}l_{2}+2\pi ip},y=e^{2ikn_{1}l_{1}+2\pi iq},k=\frac{\log x-2\pi ip}{2in_{2}l_{2}},\label{eq:xyk}
\end{equation}
where $p$ and $q$ are integers and get

\begin{equation}
\frac{\partial r_{\mathrm{num}}}{\partial k}=r_{1}n_{2}l_{2}xr_{2}+y\left[r_{2}n_{1}l_{1}+\left(n_{1}l_{1}+n_{2}l_{2}\right)x\right]=0.
\end{equation}
We express $x$ as follows 
\begin{equation}
x=-\frac{yr_{2}n_{1}l_{1}}{r_{1}n_{2}l_{2}r_{2}+y\left(n_{1}l_{1}+n_{2}l_{2}\right)},
\end{equation}
and substitute $x$ in Eq. $\eqref{eq:r_num}$ to obtain
\begin{equation}
r_{1}\left(-\frac{yr_{2}n_{1}l_{1}}{r_{1}n_{2}l_{2}r_{2}+y\left(n_{1}l_{1}+n_{2}l_{2}\right)}r_{2}+1\right)+y\left(r_{2}-\frac{yr_{2}n_{1}l_{1}}{r_{1}n_{2}l_{2}r_{2}+y\left(n_{1}l_{1}+n_{2}l_{2}\right)}\right)=0.
\end{equation}
In this way we get explicit expressions for $x$ and $y$

\begin{align}
x= & r_{1}\frac{l_{1}n_{1}r_{2}^{2}-l_{1}n_{1}-l_{2}n_{2}r_{2}^{2}-l_{2}n_{2}\text{\ensuremath{\mp\sqrt{\left(l_{1}n_{1}r_{2}^{2}+l_{1}n_{1}-l_{2}n_{2}r_{2}^{2}+l_{2}n_{2}\right)^{2}-4l_{1}^{2}n_{1}^{2}r_{2}^{2}}}}}{2l_{2}n_{2}r_{2}},\label{eq:x}\\
y= & \frac{-l_{1}n_{1}r_{2}^{2}-l_{1}n_{1}+l_{2}n_{2}r_{2}^{2}-l_{2}n_{2}\pm\sqrt{\left(l_{1}n_{1}r_{2}^{2}+l_{1}n_{1}-l_{2}n_{2}r_{2}^{2}+l_{2}n_{2}\right)^{2}-4l_{1}^{2}n_{1}^{2}r_{2}^{2}}}{2l_{1}n_{1}r_{2}}.
\end{align}
Now we formulate a relation between $x$ and $y$ by expressing $k$
with $x$ and $y$ using Eq. (\ref{eq:xyk}) and equating

\[
\frac{\log x-2\pi ip}{n_{2}l_{2}}n_{1}l_{1}=\log y-2\pi iq,
\]
\[
\log x^{\frac{n_{1}l_{1}}{n_{2}l_{2}}}-2\pi ip\frac{n_{1}l_{1}}{n_{2}l_{2}}=\log y-2\pi iq,
\]
\begin{equation}
ye^{2\pi ip\frac{n_{1}l_{1}}{n_{2}l_{2}}}=x^{\frac{n_{1}l_{1}}{n_{2}l_{2}}},
\end{equation}
which is a general EP equation for $p$ and the $q$ is not expressed
in the equation. To determine $q$ we can calculate $k$ from $x$
and $y$ and equate them. Note that this calculation is valid also
in the strong-coupling regime. Finally, we get

\begin{gather}
\left(\frac{-n_{1}r_{2}^{2}-n_{1}+\frac{l_{2}}{l_{1}}n_{2}r_{2}^{2}-\frac{l_{2}}{l_{1}}n_{2}\pm\sqrt{\left(n_{1}r_{2}^{2}+n_{1}-\frac{l_{2}}{l_{1}}n_{2}r_{2}^{2}+\frac{l_{2}}{l_{1}}n_{2}\right)^{2}-4n_{1}^{2}r_{2}^{2}}}{2n_{1}r_{2}}\right)^{\frac{n_{1}l_{1}}{n_{2}l_{2}}}=\nonumber\\
r_{1}\frac{\frac{l_{1}}{l_{2}}n_{1}r_{2}^{2}-\frac{l_{1}}{l_{2}}n_{1}-n_{2}r_{2}^{2}-n_{2}\text{\ensuremath{\mp\sqrt{\left(\frac{l_{1}}{l_{2}}n_{1}r_{2}^{2}+\frac{l_{1}}{l_{2}}n_{1}-n_{2}r_{2}^{2}+n_{2}\right)^{2}-4\left(\frac{l_{1}}{l_{2}}\right)^{2}n_{1}^{2}r_{2}^{2}}}}}{2n_{2}r_{2}}e^{2\pi ip\frac{n_{1}l_{1}}{n_{2}l_{2}}}.\label{eq:EP_equation}
\end{gather}
 Note that when the expression in the square root vanishes we have a meeting of two CPA EPs.
\subsection{Results}

\begin{figure}
\begin{centering}
\includegraphics[scale=0.8]{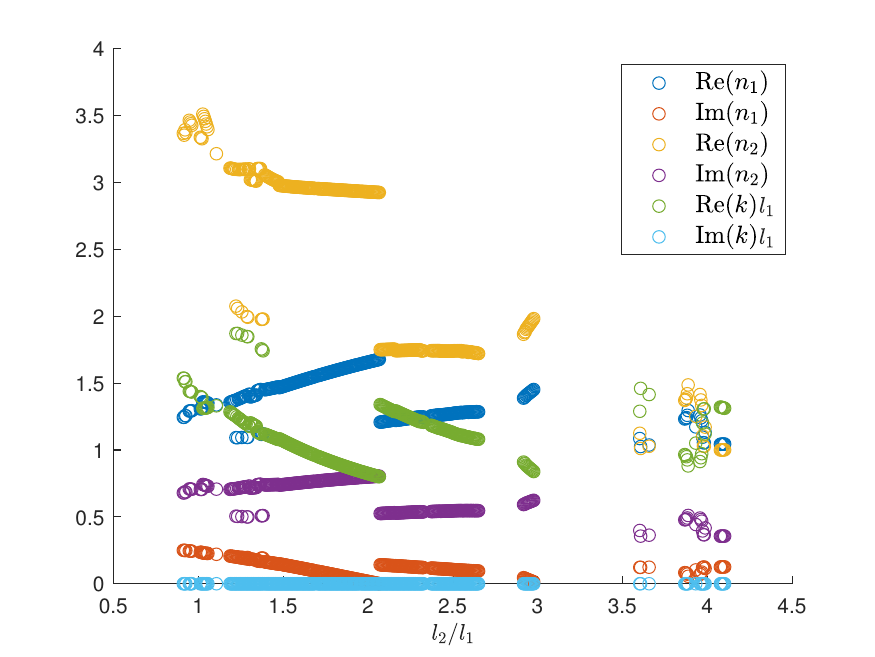}
\par\end{centering}
\caption{Coherent Perfect Absorption Exceptional Points with real $k\mathrm{s}$
as a function of $l_{2}/l_{1}$ for the two slab and perfect mirror
setup (Fig. 1 (a)). The discontinuities in the solutions arise in
some cases since we imposed $\mathrm{Im}\left(n_{1}\right),\mathrm{Im}\left(n_{2}\right)\protect\geq0$
and $\mathrm{Im\left(k\right)=0},$ and in other cases can be remedied
by tracking the solutions. \label{fig:CPA EPs}}
\end{figure}

\begin{figure}
\begin{centering}
\subfloat[]{\begin{centering}
\includegraphics[width=8cm]{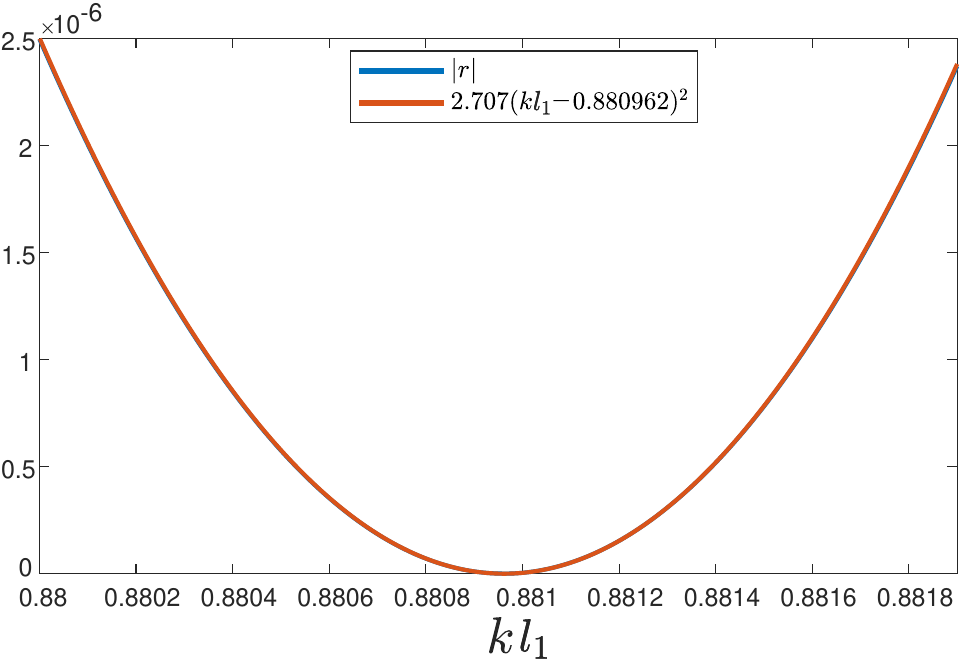}
\par\end{centering}
}\subfloat[]{\begin{centering}
\includegraphics[width=8cm]{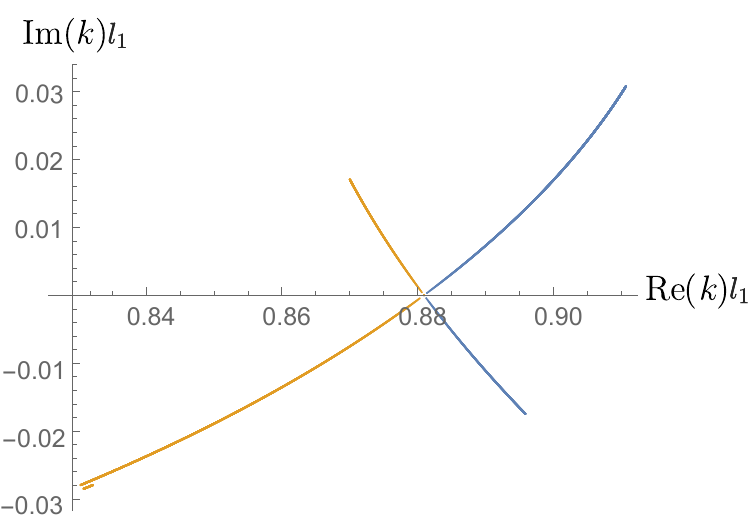}
\par\end{centering}
}
\par\end{centering}
\caption{$\left|r\right|$ as a function of $k$ close to a CPA EP (a) and
meeting of two CPAs at an EP with real $k$ $\left(b\right)$ for
the two slab setup with \label{fig:comparison} $k=0.88095,\,\,n_{1}=1.2951+0.055764i,\,\,n_{2}=1.4867+0.51067i,\,l_{1}=1,\,\,l_{2\,}=3.885.$}
\end{figure}

We first searched for a CPA EP with a real $k$ for the two slab setup
 with $n_{1}l_{1}=n_{2}l_{2},$
which means that $x=y,$ by imposing vanishing of $r_{\mathrm{num}}$
and $\partial r_{\mathrm{num}}/\partial k$ and real $k$ and obtained
agreement with our analytic results expressed using a quadratic polynomial
solution. This substantiates that the two equations enable us to calculate
an EP. We then searched for CPA EPs for $p=2$ and real $k=\frac{\log x-2\pi ip}{2in_{2}l_{2}},$
where $x$ is defined in Eq. (\ref{eq:x}) (plus solution), as we
varied $l_{2}/l_{1}$ using Eq. (\ref{eq:EP_equation}), imposing
$\mathrm{Im}\left(n_{1}\right),\mathrm{Im}\left(n_{2}\right)\geq0,$
and obtained many solutions (see Fig. \ref{fig:CPA EPs}) from which
we chose
\begin{gather}
x_{\mathrm{CPA}\,\mathrm{EP}}=-0.022158-0.020718i,\,\,\,y_{\mathrm{CPA}\,\mathrm{EP}}=-0.59158+0.68679i,\nonumber\\
k_{\mathrm{CPA}\,\mathrm{EP}}=0.88095,\,\,\,n_{1\,\mathrm{CPA}\,\mathrm{EP}}=1.2951+0.055764i,\,\,n_{2\,\mathrm{CPA}\,\mathrm{EP}}=1.4867+0.51067i,\,\nonumber\\
l_{1}=1,\,\,\,l_{2\,\mathrm{CPA}\,\mathrm{EP}}=3.8850.\nonumber
\end{gather}
In Fig. \ref{fig:comparison} we present $\left|r\right|$ as a function
of $k$ close to the CPA EP $\left(a\right)$ and the meeting of two
CPAs at the EP $\left(b\right)$. Clearly, there is a very good agreement
with the expected form of $\left(\omega-\omega_{n}\right)^{2}$ close
to an EP \cite{sweeney2019perfectly} and the EP is exact. 

To calculate an EP in the two slab setup with the Bragg mirrors we
used
\[
N_{1}=5,\,\,\,N_{2}=7,\,\,\,n_{3}=n_{5}=1.9,\,\,\,n_{4}=n_{6}=1.5,\,\,p=2,\,\,\,l_{1}=1,\,\,\,l_{2}=1.5,
\]
and obtained (see Fig. 1 (c))
\begin{gather}
n_{1\,\mathrm{EP}}=1.083+0.005i,\,\,\,n_{2\,\mathrm{EP}}=2.17+0.107i,\nonumber\\
r_{1\,\mathrm{EP}}=0.84+0.000687i,\,\,\,r_{2\,\mathrm{EP}}=0.964+0.0015i,\,\,\,k_{\mathrm{EP}}=1.4475.\nonumber
\end{gather}

Finally, to examine conversion of signals, we calculated system parameters
for a standard CPA by imposing vanishing of $r_{\mathrm{num}}$ and
real $k$ and obtained
\[
l_{1}=1,\,\,l_{2}=1.5,\,\,n_{1}=1.533,\,\,n_{2}=1.4+0.3i,\,\,k_{1}=1.2772.
\]

\section{Time Domain Calculations}

\subsection{Virtual CPA EP}

Here we calculate the temporal response by convolving the reflection
coefficient with the input. We first incorporate only the numerator,
which is associated with no scattering and then examine the effect
of the denominator.

To that end, we decompose $A_{2}/A_{1}$ into $f_{1},f_{2},f_{3}$

\begin{gather}
f_{1}=e^{-2ikl_{1}},\,\,\,f_{2}=r_{1}\left(e^{2ikl_{2}}r_{2}+1\right)+e^{2ikl_{1}}\left(r_{2}+e^{2ikl_{2}}\right),\nonumber\\
f_{3}=\frac{1}{1+e^{2ikl_{2}}r_{2}+e^{2ikl_{1}}\left(r_{2}+e^{2ikl_{2}}\right)r_{1}},\nonumber
\end{gather}
where $f_{1}$ can be omitted when calculating $r$.

We note that $f_{3}$ is a sum of a geometric series where the ratio
between adjacent terms reads
\[
q=-\left[e^{2ikl_{2}}r_{2}+e^{2ikl_{1}}\left(r_{2}+e^{2ikl_{2}}\right)r_{1}\right].
\]
In the time domain we have
\[
A_{2}/A_{1}\left(t\right)=f_{1}\left(t\right)*f_{2}\left(t\right)*f_{3}\left(t\right).
\]
We write $f_{2}\left(t\right)$ 
\[
f_{2}=r_{1}\delta\left(t\right)+r_{1}r_{2}\delta\left(t-2\frac{l_{2}}{c}\right)+r_{2}\delta\left(t-2\frac{l_{1}}{c}\right)+\delta\left(t-2\frac{l_{1}}{c}-2\frac{l_{2}}{c}\right),
\]
which when $l_{1}=l_{2}$ can be written as
\[
f_{2}=r_{1}\delta\left(t\right)+r_{2}\left(r_{1}+1\right)\delta\left(t-2\frac{l_{2}}{c}\right)+\delta\left(t-2\frac{l_{1}}{c}-2\frac{l_{2}}{c}\right).
\]

\subsubsection*{The temporal response for a continuous wave input}

We first show that the inputs result in no-scattering using $f_{2}$
in continuous wave (CW). We start with the input $y_{1}\left(t\right)=e^{-s_{n}t}$
when $l_{1}=l_{2}$
\begin{gather}
f_{2}*y_{\mathrm{1}}=r_{1}e^{-s_{n}t}+r_{2}\left(r_{1}+1\right)e^{-s_{n}\left(t-2\frac{l_{2}}{c}\right)}+e^{-s_{n}\left(t-4\frac{l_{2}}{c}\right)}=\nonumber\\
e^{-s_{n}t}\left[r_{1}+r_{2}\left(r_{1}+1\right)e^{s_{n}\left(2\frac{l_{2}}{c}\right)}+e^{s_{n}\left(4\frac{l_{2}}{c}\right)}\right]=\nonumber\\
e^{-s_{n}t}\left[r_{1}+r_{2}\left(r_{1}+1\right)x+x^{2}\right]=0,\nonumber
\end{gather}
which vanishes for a CPA independently of $e^{-s_{n}t}.$

For the general case $l_{1}\neq l_{2}$ we obtain
\begin{gather}
f_{2}*y_{1}=r_{1}e^{-s_{n}t}+r_{1}r_{2}e^{-s_{n}\left(t-2\frac{l_{1}}{c}\right)}+r_{2}e^{-s_{n}\left(t-2\frac{l_{2}}{c}\right)}r_{1}+e^{-s_{n}\left(t-2\frac{l_{1}}{c}-2\frac{l_{2}}{c}\right)}\nonumber\\
=e^{-s_{n}t}\left[r_{1}+r_{1}r_{2}e^{s_{n}\left(2\frac{l_{1}}{c}\right)}+r_{2}e^{s_{n}\left(2\frac{l_{2}}{c}\right)}r_{1}+e^{s_{n}\left(2\frac{l_{1}}{c}+2\frac{l_{2}}{c}\right)}\right]\nonumber\\
=e^{-s_{n}t}\left[r_{1}+r_{1}r_{2}e^{s_{n}\left(2\frac{l_{1}}{c}\right)}+r_{2}e^{s_{n}\left(2\frac{l_{2}}{c}\right)}r_{1}+e^{s_{n}\left(2\frac{l_{1}}{c}+2\frac{l_{2}}{c}\right)}\right]=0,\nonumber
\end{gather}
which also vanishes for a CPA independently of $e^{-s_{n}t}.$

Now we analyze the input $y_{2}\left(t\right)=te^{-s_{n}t}$ when $l_{1}=l_{2}$
\begin{gather}
f_{2}*y_{2}=r_{1}te^{-s_{n}t}+r_{2}\left(r_{1}+1\right)\left(t-2\frac{l_{2}}{c}\right)e^{-s_{n}\left(t-2\frac{l_{2}}{c}\right)}+\left(t-4\frac{l_{2}}{c}\right)e^{-s_{n}\left(t-4\frac{l_{2}}{c}\right)}\nonumber\\
=t\left(r_{1}e^{s_{n}t}+r_{2}\left(r_{1}+1\right)e^{-s_{n}\left(t-2\frac{l_{2}}{c}\right)}+e^{-s_{n}\left(t-4\frac{l_{2}}{c}\right)}\right)\nonumber\\
+r_{2}\left(r_{1}+1\right)\left(-2\frac{l_{2}}{c}\right)e^{-s_{n}\left(t-2\frac{l_{2}}{c}\right)}+\left(-4\frac{l_{2}}{c}\right)e^{-s_{n}\left(t-4\frac{l_{2}}{c}\right)}\nonumber\\
=t\left(r_{1}e^{-s_{n}t}+r_{2}\left(r_{1}+1\right)e^{-s_{n}\left(t-2\frac{l_{2}}{c}\right)}+e^{-s_{n}\left(t-4\frac{l_{2}}{c}\right)}\right)\nonumber\\
+e^{-s_{n}t}\left[r_{2}\left(r_{1}+1\right)\left(-2\frac{l_{2}}{c}\right)e^{s_{n}\left(2\frac{l_{2}}{c}\right)}+\left(-4\frac{l_{2}}{c}\right)e^{s_{n}\left(4\frac{l_{2}}{c}\right)}\right]\nonumber\\
=te^{-s_{n}t}\left(r_{1}+r_{2}\left(r_{1}+1\right)e^{2\frac{l_{2}}{c}}+e^{4\frac{l_{2}}{c}}\right)+e^{-s_{n}t}\left[r_{2}\left(r_{1}+1\right)\left(-2\frac{l_{2}}{c}\right)x+\left(-4\frac{l_{2}}{c}\right)x^{2}\right]\nonumber\\
=2e^{-s_{n}t}x\frac{l_{2}}{c}\left[r_{2}\left(r_{1}+1\right)+2x\right].\nonumber
\end{gather}
Since $x=-\frac{b}{2a}$ at the EP we get
\[
f_{2}*y_{\mathrm{2}}=0.
\]
For the general case $l_{1}\neq l_{2}$ we obtain 
\begin{gather}
f_{2}*y_{2}=tr_{1}e^{-s_{n}t}+r_{1}r_{2}\left(t-2\frac{l_{1}}{c}\right)e^{-s_{n}\left(t-2\frac{l_{1}}{c}\right)}+\left(t-2\frac{l_{2}}{c}\right)r_{2}e^{-s_{n}\left(t-2\frac{l_{2}}{c}\right)}r_{1}+\left(t-2\frac{l_{1}}{c}-2\frac{l_{2}}{c}\right)e^{-s_{n}\left(t-2\frac{l_{1}}{c}-2\frac{l_{2}}{c}\right)}\nonumber\\
=e^{-s_{n}t}t\left[r_{1}+r_{1}r_{2}e^{s_{n}2\frac{l_{1}}{c}}+r_{2}e^{s_{n}2\frac{l_{2}}{c}}r_{1}+e^{s_{n}\left(2\frac{l_{1}}{c}+2\frac{l_{2}}{c}\right)}\right]\nonumber\\
+e^{-s_{n}t}\left[r_{1}r_{2}\left(-2\frac{l_{1}}{c}\right)e^{s_{n}\left(2\frac{l_{1}}{c}\right)}+\left(-2\frac{l_{2}}{c}\right)r_{2}e^{s_{n}\left(2\frac{l_{2}}{c}\right)}r_{1}+\left(-2\frac{l_{1}}{c}-2\frac{l_{2}}{c}\right)e^{s_{n}\left(2\frac{l_{1}}{c}+2\frac{l_{2}}{c}\right)}\right]=0\nonumber
\end{gather}
so the derivative of the numerator with respect to $s$ needs to vanish
to satisfy $f_{2}*y_{2}$ as expected at an EP. This is in agreement
with the expected form of $\left(s-s_{n}\right)^{2}\rightarrow$ $\left(s-s_{n}\right)=0$
upon differentiation.

\subsubsection{Introducing step functions in the input}

Now we introduce step functions in the input
\[
y_{\mathrm{1u}}=e^{-s_{n}t}\left[\theta\left(t\right)-\theta\left(t-t_{1}\right)\right],
\]
\[
f_{2}*y_{\mathrm{1}u}=e^{-s_{n}t}\left[r_{1}\theta\left(t\right)+r_{1}r_{2}e^{s_{n}\left(2\frac{l_{1}}{c}\right)}\theta\left(t-2\frac{l_{1}}{c}\right)+r_{2}e^{s_{n}\left(2\frac{l_{2}}{c}\right)}r_{1}\theta\left(t-2\frac{l_{2}}{c}\right)+e^{s_{n}\left(2\frac{l_{1}}{c}+2\frac{l_{2}}{c}\right)}\theta\left(t-2\frac{l_{1}}{c}-2\frac{l_{2}}{c}\right)\right]
\]
\[
-e^{-s_{n}t}\left[r_{1}\theta\left(t-t_{1}\right)+r_{1}r_{2}e^{s_{n}\left(2\frac{l_{1}}{c}\right)}\theta\left(t-2\frac{l_{1}}{c}-t_{1}\right)+r_{2}e^{s_{n}\left(2\frac{l_{2}}{c}\right)}r_{1}\theta\left(t-2\frac{l_{2}}{c}-t_{1}\right)+e^{s_{n}\left(2\frac{l_{1}}{c}+2\frac{l_{2}}{c}\right)}\theta\left(t-2\frac{l_{1}}{c}-2\frac{l_{2}}{c}-t_{1}\right)\right].
\]
After $2\frac{l_{1}}{c}+2\frac{l_{2}}{c}<t<t_{1},t_{1}+2\frac{l_{1}}{c}+2\frac{l_{2}}{c}<t$
there is no scattering (still needs to be convoluted with $f_{3}$)
since we return to the previous expression. For $t<2\frac{l_{1}}{c}+2\frac{l_{2}}{c}$
the scattering will originate from $\theta\left(t\right)$ and for $t_{1}<t<t_{1}+2\frac{l_{1}}{c}+2\frac{l_{2}}{c}$
the scattering will originate from $-\theta\left(t-t_{1}\right)$ as expected.
For a virtual CPA $e^{-s_{n}t}$ will be large for $t_{1}<t<t_{1}+2\frac{l_{1}}{c}+2\frac{l_{2}}{c},$
so there is strong scattering.

Now we analyze
\[
y_{\mathrm{2u}}\left(t\right)=te^{-s_{n}t}\left[\theta\left(t\right)-\theta\left(t-t_{1}\right)\right].
\]
We start with $y_{\mathrm{2}u1}=te^{-s_{n}t}\theta\left(t\right)$
\[
f_{2}*y_{\mathrm{2u1}}=tr_{1}e^{-s_{n}t}\theta\left(t\right)+r_{1}r_{2}\left(t-2\frac{l_{1}}{c}\right)e^{-s_{n}\left(t-2\frac{l_{1}}{c}\right)}\theta\left(t-2\frac{l_{1}}{c}\right)
\]
\[
+\left(t-2\frac{l_{2}}{c}\right)r_{2}e^{-s_{n}\left(t-2\frac{l_{2}}{c}\right)}r_{1}\theta\left(t-2\frac{l_{2}}{c}\right)+\left(t-2\frac{l_{1}}{c}-2\frac{l_{2}}{c}\right)e^{-s_{n}\left(t-2\frac{l_{1}}{c}-2\frac{l_{2}}{c}\right)}\theta\left(t-2\frac{l_{1}}{c}-2\frac{l_{2}}{c}\right)=
\]
\[
t\left[r_{1}e^{-s_{n}t}\theta\left(t\right)+e^{-s_{n}\left(t-2\frac{l_{1}}{c}\right)}\theta\left(t-2\frac{l_{1}}{c}\right)+r_{2}e^{-s_{n}\left(t-2\frac{l_{2}}{c}\right)}r_{1}\theta\left(t-2\frac{l_{2}}{c}\right)+e^{-s_{n}\left(t-2\frac{l_{1}}{c}-2\frac{l_{2}}{c}\right)}\theta\left(t-2\frac{l_{1}}{c}-2\frac{l_{2}}{c}\right)\right]
\]
\[
+r_{1}r_{2}\left(-2\frac{l_{1}}{c}\right)e^{-s_{n}\left(t-2\frac{l_{1}}{c}\right)}\theta\left(t-2\frac{l_{1}}{c}\right)+\left(-2\frac{l_{2}}{c}\right)r_{2}e^{-s_{n}\left(t-2\frac{l_{2}}{c}\right)}r_{1}\theta\left(t-2\frac{l_{2}}{c}\right)
\]
\[
+\left(-2\frac{l_{1}}{c}-2\frac{l_{2}}{c}\right)e^{-s_{n}\left(t-2\frac{l_{1}}{c}-2\frac{l_{2}}{c}\right)}\theta\left(t-2\frac{l_{1}}{c}-2\frac{l_{2}}{c}\right).
\]
and we observe that when $t>2\frac{l_{1}}{c}+2\frac{l_{2}}{c}$ there
is no scattering (still needs to be convoluted) since it can be written
in the CW form. 

Now we analyze $y_{\mathrm{2\,}u2}=-te^{-s_{n}t}\theta\left(t-t_{1}\right)$
\[
f_{2}*y_{\mathrm{2\,}u2}=-\left[tr_{1}e^{-s_{n}t}\theta\left(t-t_{1}\right)+r_{1}r_{2}\left(t-2\frac{l_{1}}{c}-t_{1}\right)e^{-s_{n}\left(t-2\frac{l_{1}}{c}\right)}\theta\left(t-2\frac{l_{1}}{c}-t_{1}\right)\right.
\]
\[
\left.+\left(t-2\frac{l_{2}}{c}\right)r_{2}e^{-s_{n}\left(t-2\frac{l_{2}}{c}\right)}r_{1}\theta\left(t-2\frac{l_{2}}{c}-t_{1}\right)+\left(t-2\frac{l_{1}}{c}-2\frac{l_{2}}{c}\right)e^{-s_{n}\left(t-2\frac{l_{1}}{c}-2\frac{l_{2}}{c}\right)}\theta\left(t-2\frac{l_{1}}{c}-2\frac{l_{2}}{c}-t_{1}\right)\right]
\]
and observe that from $t=t_{1}+\frac{2\left(l_{1}+l_{2}\right)}{c}$
there is no scattering.

\subsubsection{Incorporating the denominator of the reflection coefficient in the
temporal response calculation}

Now we convolve $f_{2}*y_{1u}$ with $f_{1}$ in order to develop
intuition for the effect of an exponent in frequency 
\[
f_{1}*f_{2}*y_{1u}=
\]
\[
=e^{-s_{n}t}\left[r_{1}e^{-s_{n}\frac{2l_{1}}{c}}\theta\left(t+2\frac{l_{1}}{c}\right)+r_{1}r_{2}e^{s_{n}\left(2\frac{l_{1}}{c}-2\frac{l_{1}}{c}\right)}\theta\left(t-2\frac{l_{1}}{c}+2\frac{l_{1}}{c}\right)+r_{2}e^{s_{n}\left(2\frac{l_{2}}{c}-2\frac{l_{1}}{c}\right)}r_{1}\theta\left(t-2\frac{l_{2}}{c}+2\frac{l_{1}}{c}\right)\right.
\]
\[
\left.+e^{s_{n}\left(2\frac{l_{1}}{c}+2\frac{l_{2}}{c}-2\frac{l_{1}}{c}\right)}\theta\left(t-2\frac{l_{1}}{c}-2\frac{l_{2}}{c}+2\frac{l_{1}}{c}\right)\right]
\]
\[
-e^{-s_{n}t}\left[r_{1}e^{-s_{n}\frac{2l_{1}}{c}}\theta\left(t-t_{1}+2\frac{l_{1}}{c}\right)+r_{1}r_{2}e^{s_{n}\left(2\frac{l_{1}}{c}-2\frac{l_{1}}{c}\right)}\theta\left(t-2\frac{l_{1}}{c}-t_{1}+2\frac{l_{1}}{c}\right)\right.
\]
\[
\left.+r_{2}e^{s_{n}\left(2\frac{l_{2}}{c}-2\frac{l_{1}}{c}\right)}r_{1}\theta\left(t-2\frac{l_{2}}{c}-t_{1}+2\frac{l_{1}}{c}\right)+e^{s_{n}\left(2\frac{l_{1}}{c}+2\frac{l_{2}}{c}-2\frac{l_{1}}{c}\right)}\theta\left(t-2\frac{l_{1}}{c}-2\frac{l_{2}}{c}-t_{1}+2\frac{l_{1}}{c}\right)\right]
\]
\[=e^{-s_{n}\left(t+2\frac{l_{1}}{c}\right)}\left[r_{1}\theta\left(t+2\frac{l_{1}}{c}\right)+r_{1}r_{2}e^{s_{n}\left(2\frac{l_{1}}{c}\right)}\theta\left(t-2\frac{l_{1}}{c}+2\frac{l_{1}}{c}\right)+r_{2}e^{s_{n}\left(2\frac{l_{2}}{c}\right)}r_{1}\theta\left(t-2\frac{l_{2}}{c}+2\frac{l_{1}}{c}\right)\right.
\]
\[
\left.+e^{s_{n}\left(2\frac{l_{1}}{c}+2\frac{l_{2}}{c}\right)}\theta\left(t-2\frac{l_{1}}{c}-2\frac{l_{2}}{c}+2\frac{l_{1}}{c}\right)\right]
\]
\[
-e^{-s_{n}\left(t+2\frac{l_{1}}{c}\right)}\left[r_{1}\theta\left(t-t_{1}+2\frac{l_{1}}{c}\right)+r_{1}r_{2}e^{s_{n}\left(2\frac{l_{1}}{c}\right)}\theta\left(t-2\frac{l_{1}}{c}-t_{1}+2\frac{l_{1}}{c}\right)\right.
\]
\[
\left.+r_{2}e^{s_{n}\left(2\frac{l_{2}}{c}\right)}r_{1}\theta\left(t-2\frac{l_{2}}{c}-t_{1}+2\frac{l_{1}}{c}\right)+e^{s_{n}\left(2\frac{l_{1}}{c}+2\frac{l_{2}}{c}\right)}\theta\left(t-2\frac{l_{1}}{c}-2\frac{l_{2}}{c}-t_{1}+2\frac{l_{1}}{c}\right)\right].
\]
It can be seen that $f_{1}$ just shifted the times in which there is no scattering. It
also introduced a phase that for a complex $k$ can increase the amplitude.

Thus, multiplication by an exponent in frequency or convolution with
a delta function in time results in a time delay 
\[
\delta\left(t-t_{n}\right)*g\rightarrow e^{s_{n}t_{1}}g\left(t-t_{n}\right),
\]
which shifts the times in which there is no scattering and introduces
an exponent factor that might change the amplitude.

Similarly for the second input, we write $y_{\mathrm{2}u1}=te^{-s_{n}t}\theta\left(t\right)$
$f_{1}:t\rightarrow t-2\frac{l_{1}}{c}$
\[
f_{1}*f_{2}*y_{\mathrm{2u1}}=
\]
\[
\left(t-2\frac{l_{1}}{c}\right)\left[r_{1}e^{-s_{n}\left(t-2\frac{l_{1}}{c}\right)}\theta\left(t-2\frac{l_{1}}{c}\right)+e^{-s_{n}\left(\left(t-2\frac{l_{1}}{c}\right)-2\frac{l_{1}}{c}\right)}\theta\left(\left(t-2\frac{l_{1}}{c}\right)-2\frac{l_{1}}{c}\right)\right.
\]
\[
\left.+r_{2}e^{-s_{n}\left(\left(t-2\frac{l_{1}}{c}\right)-2\frac{l_{2}}{c}\right)}r_{1}\theta\left(\left(t-2\frac{l_{1}}{c}\right)-2\frac{l_{2}}{c}\right)+e^{-s_{n}\left(\left(t-2\frac{l_{1}}{c}\right)-2\frac{l_{1}}{c}-2\frac{l_{2}}{c}\right)}\theta\left(\left(t-2\frac{l_{1}}{c}\right)-2\frac{l_{1}}{c}-2\frac{l_{2}}{c}\right)\right]
\]
\[
+r_{1}r_{2}\left(-2\frac{l_{1}}{c}\right)e^{-s_{n}\left(\left(t-2\frac{l_{1}}{c}\right)-2\frac{l_{1}}{c}\right)}\theta\left(t-2\frac{l_{1}}{c}\right)+\left(-2\frac{l_{2}}{c}\right)r_{2}e^{-s_{n}\left(\left(t-2\frac{l_{1}}{c}\right)-2\frac{l_{2}}{c}\right)}r_{1}\theta\left(\left(t-2\frac{l_{1}}{c}\right)-2\frac{l_{2}}{c}\right)
\]
\[
+\left(-2\frac{l_{1}}{c}-2\frac{l_{2}}{c}\right)e^{-s_{n}\left(\left(t-2\frac{l_{1}}{c}\right)-2\frac{l_{1}}{c}-2\frac{l_{2}}{c}\right)}\theta\left(\left(t-2\frac{l_{1}}{c}\right)-2\frac{l_{1}}{c}-2\frac{l_{2}}{c}\right).
\]
We can see that after the first round trip we get a contribution of
the same type $t\cdot\mathrm{numerator}$. At $t=2\frac{l_{1}}{c}$
we get an amplitude that is smaller since it's from earlier times.
Conclusion: compared to the input, there is less reflection. 

For a virtual CPA we get
\[
e^{s_{n}t_{1}}g\left(t-t_{n}\right)=e^{i\omega_{n}t_{1}}g\left(t-t_{n}\right)=e^{i\left(\mathrm{Re}(\omega_{n})+i\mathrm{Im}(\omega_{n})\right)t_{1}}g\left(t-t_{n}\right),
\]
and since $\mathrm{Im}(\omega_{n})>0,$ $e^{-\mathrm{Im}(\omega_{n})}<1,$
which makes sense since it's the signal at earlier times that is smaller. 

Similarly, the series in $f_{3}$ can be truncated at e.g., $N=7$
for our EP parameters and each term just shifts the times in which
there is no scattering (introduces time delay). Here, there are also reflectivities which reduce the amplitude of the scattered field at later times. Clearly, high-Q cavities have larger reflectivities, which means that the equilibration time and the reflection will be larger
\[
f_{3}=\sum_{n=1}^{N}\left[-\left(e^{2ikl_{2}}r_{2}+e^{2ikl_{1}}\left(r_{2}+e^{2ikl_{2}}\right)r_{1}\right)\right]^{n}
\]
\[
f_{3}=1-\left(r_{2}e^{2ikl_{2}}+r_{1}r_{2}e^{2ikl_{1}}+r_{1}e^{2ik\left(l_{1}+l_{2}\right)}\right)+\left(r_{2}e^{2ikl_{2}}+r_{1}r_{2}e^{2ikl_{1}}+r_{1}e^{2ik\left(l_{1}+l_{2}\right)}\right)^{2}+...
\]
\[
f_{3}=1-\left(r_{2}e^{2ikl_{2}}+r_{1}r_{2}e^{2ikl_{1}}+r_{1}e^{2ik\left(l_{1}+l_{2}\right)}\right)
\]
\[
+\left(r_{2}^{2}e^{4ikl_{2}}+r_{1}^{2}r_{2}^{2}e^{4ikl_{1}}+r_{1}^{2}e^{4ik\left(l_{1}+l_{2}\right)}+2r_{1}r_{2}^{2}e^{2ik\left(l_{1}+l_{2}\right)}+2r_{1}r_{2}e^{2ik\left(l_{1}+2l_{2}\right)}+2r_{1}^{2}r_{2}e^{2ik\left(2l_{1}+l_{2}\right)}\right)+...
\]
\[
f_{3}\left(t\right)=1-\left[r_{2}\delta\left(t-2l_{2}\right)+r_{1}r_{2}\delta\left(t-2l_{1}\right)+r_{1}\delta\left(t-2\left(l_{1}+l_{2}\right)\right)\right]+
\]
\[
\left[r_{2}^{2}\delta\left(t-4l_{2}\right)+r_{1}^{2}r_{2}^{2}\delta\left(t-4l_{1}\right)+r_{1}^{2}\delta\left(t-4l_{1}-4l_{2}\right)+2r_{1}r_{2}^{2}\delta\left(t-2\left(l_{1}+l_{2}\right)\right)+2r_{1}r_{2}\delta\left(t-2\left(l_{1}+2l_{2}\right)\right)\right.
\]
\[
\left.+2r_{1}^{2}r_{2}\delta\left(t-4\left(l_{1}+l_{2}\right)\right)\right]\]
For $N=2$ two round trips in one cavity is the shortest delay and
two round trips in both cavities is the longest delay. Thus, the maximal
delay is $\left(N+1\right)\cdot\mathrm{round\,trip}\,\mathrm{time}$.
For our EP parameters, after approximately 5 or 7 roundtrips, there
will be negligible scattering (high orders of reflection coefficients).
In a high-Q cavity this time will be longer. Then, after $t_{1},$ $-\theta\left(t\right)$
will kick in so there will be scattering and then after 5 or 7 roundtrips,
there will be negligible scattering. We assumed $t_{1}>\mathrm{equilbration\,time}.$
To approximate how many terms are needed we can take the largest reflection
coefficient and require $r_{2}^{N}<0.02,$ which for $r_{2}=0.57,$ results
in $N=7.$
\label{SM_energy}

\begin{figure}
\subfloat[]{\includegraphics[scale=0.72]{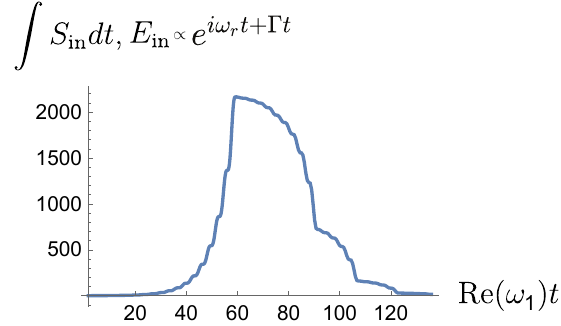}
}\subfloat[]{\includegraphics[scale=0.77]{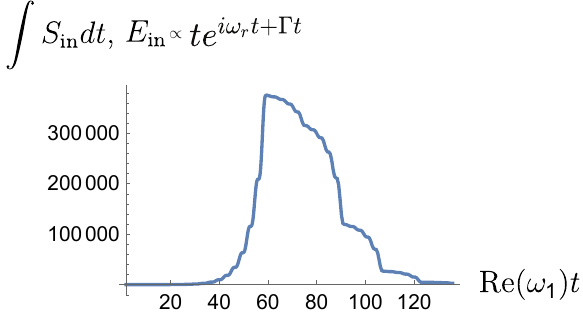}
}

\caption{Input and scattered energy as a function of time for the inputs of 
$e^{i\omega_{r}t+\Gamma t}$ (a) and $te^{i\omega_{r}t+\Gamma t}$ (b) multiplied by
$\left(\theta\left(t-t_{1}\right)-\theta\left(t-t_{2}\right)\right).$ All the
stored energy is released after the input signal stops. In addition,
the temporarily absorbed energy for the input of $te^{i\omega_{r}t+\Gamma t}$
is $\sim$1000 times larger. \label{fig:Power}}
\end{figure}

\subsection{Time Domain calculations for a real CPA and real CPA EP using a novel method to perform numerical Inverse Fourier Transform}

First, we present the Fourier Transform of the inputs multiplied by step functions. 
\[
\mathcal{F}\left\{ (\theta(t)-\theta(t-t_{1}))\exp(-i\text{\ensuremath{k_{\mathrm{EP}}}}t)\right\} =\frac{ie^{-t_{1}ik_{\mathrm{EP}}}\left(e^{t_{1}i\text{\ensuremath{k_{\mathrm{EP}}}}}-e^{t_{1}ik}\right)}{\sqrt{2\pi}(k-\text{\ensuremath{k_{\mathrm{EP}}}})},
\]
\[
\mathcal{F}\left\{ (\theta(t)-\theta(t-t_{1}))t\exp(-i\text{\ensuremath{k_{\mathrm{EP}}}}t)\right\} =\frac{e^{-t_{1}ik_{\mathrm{EP}}}\left(e^{t_{1}ik}(-t_{1}k+t_{1}i\text{\ensuremath{k_{\mathrm{EP}}}}+1)-e^{t_{1}i\text{\ensuremath{k_{\mathrm{EP}}}}}\right)}{\sqrt{2\pi}(k-k_{\mathrm{EP}})^{2}}.
\]

A direct IFT calculation has several issues. First, at negative frequencies
the incoming and scattered fields switch roles and we get $r\left(-\omega\right)=1/r\left(\omega\right),$
which results in divergences for a real CPA EP and CPA. Second, at
a CPA EP there are expressions in $r\left(k\right)\cdot\mathrm{input}\left(k\right)$
of the forms $\frac{(k-k_{\mathrm{EP}})^{2}}{(k-k_{\mathrm{EP}})},\frac{(k-k_{\mathrm{EP}})^{2}}{(k-k_{\mathrm{EP}})^{2}},$
which have numerical divergences. Third, $\underset{\omega\rightarrow\infty}{\lim}r\left(\omega\right)=\mathrm{constant}$
and $\theta\left(t\right)$ decays slowly at high frequencies and
when Inverse Fourier transforming for large times it becomes computationally
challenging since the integrand strongly oscillates and the integration
range is large. Fourth, numerical FT cannot treat the delta function
that results from switching off the input.

In order to alleviate these issues we have employed several approaches.
First, since for a real input denoted by $A,$ satisfied when $A\left(-\omega\right)=A^{*}\left(\omega\right),$
we expect a real output, we require that the transfer function that
relates between input to output will satisfy $H\left(-\omega\right)=H^{*}\left(\omega\right)$
\[
\left(A\left(\omega\right)H\left(\omega\right)\right)^{*}=\left(A^{*}\left(\omega\right)H^{*}\left(\omega\right)\right)=A\left(-\omega\right)H\left(-\omega\right),
\]
and in this way we rederive $H\left(-\omega\right).$ The complex
conjugation of $H\left(\omega\right)$ results in a change only in
$\mathrm{Im}\left(n_{i}\right),$ where $n_{i}$ is the refractive
index, which means that only when $\mathrm{Im}\left(n_{i}\right)=0$
the standard IFT calculation is accurate. Using this approach we Inverse
Fourier Transformed $e^{in_{i}k}$ and obtained a Lorentzian in time
$\frac{\sqrt{\frac{2}{\pi}}\mathrm{Im}\left(n_{i}\right)}{\text{\ensuremath{\mathrm{Im}\left(n_{i}\right)}}^{2}+(\mathrm{Re}\left(n_{i}\right)-t)^{2}}$
instead of a delta function. 

Second, we smoothed $r\left(k\right)\cdot\mathrm{input}\left(k\right)$
analytically close to $k_{\mathrm{EP}}$ by taking the limits to avoid
the numerical divergences. Third, we exchanged in the inputs $\theta\left(t\right)\rightarrow\left(\tanh\left(t\right)+1\right)/2$
and obtained the following FTs for the inputs with $e^{i\omega_{n}t},\,\,\,te^{i\omega_{n}t}$
\begin{equation}
\mathcal{F}\left\{ \frac{1}{2}\left(\tanh\left(t\right)-\mathrm{tanh}\left(t-t_{1}\right)\right)e^{-i\omega_{n}t}\right\} =\frac{1}{2}i\sqrt{\frac{\pi}{2}}\text{csch}\left(\frac{1}{2}\pi(k-k_{\mathrm{EP}})\right)-\frac{1}{2}i\sqrt{\frac{\pi}{2}}e^{it_{1}(k-\text{\ensuremath{k_{\mathrm{EP}}}})}\text{csch}\left(\frac{1}{2}\pi(k-k_{\mathrm{EP}})\right),
\end{equation}
\[
\mathcal{F}\left\{ \frac{1}{2}\left(\tanh\left(t\right)-\mathrm{tanh}\left(t-t_{1}\right)\right)te^{-i\omega_{n}t}\right\} =\frac{1}{2}\left\{ -\frac{\pi^{3/2}}{2\sqrt{2}}\mathrm{coth}\left(\frac{1}{2}\pi\left(\omega-\omega_{n}\right)\right)\mathrm{csch}\left(\frac{1}{2}\pi\left(\omega-\omega_{n}\right)\right)\right.
\]
\begin{equation}
\left.+e^{it_{1}\left(\omega-\omega_{n}\right)}\left[-i\sqrt{\frac{\pi}{2}}t_{1}\mathrm{csch}\left(\frac{1}{2}\pi\left(\omega-\omega_{n}\right)\right)+\frac{\pi^{3/2}}{2\sqrt{2}}e^{it_{1}\left(\omega-\omega_{n}\right)}\mathrm{coth}\left(\frac{1}{2}\pi\left(\omega-\omega_{n}\right)\right)\mathrm{csch}\left(\frac{1}{2}\pi\left(\omega-\omega_{n}\right)\right)\right]\right\} ,
\end{equation}
which decay exponentially away from $k_{\mathrm{EP}}$ and have the
same behavior close to $k_{\mathrm{EP}}$
\[
\lim_{k\rightarrow k_{\mathrm{EP}}}\text{csch}\left(\frac{1}{2}\pi(k-k_{\mathrm{EP}})\right)=\frac{1}{x},x\equiv\frac{1}{2}\pi\left(k-k_{\mathrm{EP}}\right),
\]
\[
\lim_{k\rightarrow k_{\mathrm{EP}}}\text{csch}\left(\frac{1}{2}\pi(k-k_{\mathrm{EP}})\right)\mathrm{coth}\left(\frac{1}{2}\pi\left(\omega-k_{\mathrm{EP}}\right)\right)=\frac{a}{x}+\frac{b}{x^{2}}.
\]
 Fourth, the FT of $e^{it_{1}(k-\text{\ensuremath{k_{\mathrm{EP}}}})}$
was treated analytically by performing a discrete time shift. Thus,
we performed the IFT of $r\left(k\right)\cdot\mathrm{inp}_{1/2}\left(k\right),$
where $\mathrm{inp}_{1},\mathrm{inp}_{2}$ denote the first and second
inputs. 

Now we analyze qualitatively the temporal responses to the finite-time
inputs at a CPA EP and write
\begin{gather}
\lim_{k\rightarrow k_{\mathrm{EP}}}\mathcal{F}\left(\mathrm{inp}_{1}\right)\cdot r\propto\left(k-k_{\mathrm{EP}}\right),
\nonumber\\
\lim_{k\rightarrow k_{\mathrm{EP}}}\mathcal{F}\left(\mathrm{inp}_{2}\right)\cdot r=\mathrm{constant,}\nonumber
\end{gather}
which means that the first and second inputs are expected to result
in negligible oscillations and oscillation that decays in time, respectively.
Thus, the response at a CPA EP to $te^{i\omega_{n}t}$ is similar
to the response of a CPA to $e^{i\omega_{n}t}.$
Interestingly, in the FT of the input $t\exp(i\omega t)$ in Eq. (14),
the 1st term that is multiplied by $\exp(it_{1}\omega)$ is also multiplied
by $t_{1}.$ Since the FT of $\exp(i\omega t_{1})$ results in a convolution
with $\delta\left(t-t_{1}\right)$ and hence a shift in time of $t_{1},$
it means that the response will have a term that is multiplied by
$t_{1}$ after that $t_{1}$ time has passed, and thus this term might
be associated with accumulation of energy. At the $k_{\mathrm{EP}}$
limit this term behaves as $t_{1}\mathrm{csch}(\pi(k-k_{\mathrm{EP}})/2))\rightarrow t_{1}/(k-k_{\mathrm{EP}})$.
Hence, $r\cdot\mathrm{inp_{2}}$ at a CPA EP will be proportional
to $t_{1}(k-k_{\mathrm{EP}})^{2}/(k-k_{\mathrm{EP}})=0$ and at a
CPA to $t_{1}(k-k_{\mathrm{EP}})^{2}/(k-k_{\mathrm{EP}})^{2}=t_{1},$
which means that there is a small accumulation of energy at a CPA
EP (due to the side lobes of $r\cdot\mathrm{inp_{2})}$ and a large
accumulation of energy at a CPA when the input is $t\exp(i\omega t).$
Importantly, this analysis of energy accumulation appears to be independent
of $r$ and thus this input may be fundamental in light matter interactions.

Finally, we analyzed the temporal response for a finite-time input
at a standard CPA. To that end, we multiplied the two inputs by $r$
at a CPA and obtained
\[
\frac{k-k_{\mathrm{CPA}}}{k-k_{\mathrm{CPA}}}=\mathrm{constant},\,\,\,\frac{k-k_{\mathrm{CPA}}}{\left(k-k_{\mathrm{CPA}}\right)^{2}}=\frac{1}{k-k_{\mathrm{CPA}}},
\]
where the second expression diverges and implies a step function multiplied
by an oscillating signal in time as expected. We now calculate the
IFT of the second input multiplied by $r$ at a CPA 
\[
\mathcal{F}^{-1}\left\{ r_{\mathrm{CPA}}\cdot\mathrm{inp}_{2}\right\} =\frac{1}{\sqrt{2\pi}}\left[\intop_{0}^{k_{\mathrm{CPA}}-0.01}r\cdot\mathrm{inp}_{2}e^{ikt}+\intop_{k_{\mathrm{CPA}}-0.01}^{k_{\mathrm{CPA}}+0.01}\frac{1}{\left(k-k_{\mathrm{CPA}}\right)}\left(a+be^{it_{1}(k-\text{\ensuremath{k_{\mathrm{CPA}}}})}\right)e^{ikt}+\intop_{k_{\mathrm{CPA}}+0.01}^{\infty}r\cdot\mathrm{inp}_{2}e^{ikt}\right]
\]
\[
=\frac{1}{\sqrt{2\pi}}\left[\intop_{0}^{k_{\mathrm{CPA}}-0.1}r\cdot\mathrm{inp}_{2}e^{ikt}+\intop_{-\infty}^{\infty}\mathrm{window}\left(50\left(k-k_{\mathrm{CPA}}\right)\right)\frac{1}{\left(k-k_{\mathrm{CPA}}\right)}\left(a+be^{it_{1}(k-\text{\ensuremath{k_{\mathrm{CPA}}}})}\right)e^{ikt}+\intop_{k_{\mathrm{CPA}}+0.1}^{\infty}r\cdot\mathrm{inp}_{2}e^{ikt}\right].
\]
\[
\frac{1}{\sqrt{2\pi}}\intop_{-\infty}^{\infty}\mathrm{window}\left(50\left(k-k_{\mathrm{CPA}}\right)\right)\frac{1}{\left(k-k_{\mathrm{CPA}}\right)}\frac{1}{2\sqrt{2}}\pi^{3/2}\sqrt{2\pi}\left(-1+e^{it_{1}(k-\text{\ensuremath{k_{\mathrm{CPA}}}})}\right)e^{ikt}
\]
\[
=\left(e^{-itk_{\mathrm{CPA}}}\sqrt{\frac{2}{\pi}}\mathrm{sinc}\left(\frac{t}{100}\right)*e^{-itk_{\mathrm{CPA}}}\left(-i\right)\sqrt{\frac{\pi}{2}}\mathrm{sign}\left(t\right)\right)*\frac{1}{2\sqrt{2}}\pi^{3/2}\sqrt{2\pi}\left(-\delta\left(t\right)+e^{-it_{1}\text{\ensuremath{k_{\mathrm{CPA}}}}}\delta\left(t-t_{1}\right)\right)
\]
\[
=\left(-2i\mathrm{Si}\left(\frac{t}{100}\right)e^{-itk_{\mathrm{CPA}}}\right)*\frac{\pi^{2}}{2}\left(-\delta\left(t\right)+e^{-it_{1}\text{\ensuremath{k_{\mathrm{CPA}}}}}\delta\left(t-t_{1}\right)\right)
\]
\[
=-i\pi^{2}e^{-itk_{\mathrm{CPA}}}\left[-\mathrm{Si}\left(\frac{t}{100}\right)+\mathrm{Si}\left(\frac{t-t_{1}}{100}\right)\right]
\]
\[
=i\pi^{2}e^{-itk_{\mathrm{CPA}}}\left[\mathrm{Si}\left(\frac{t}{100}\right)-\mathrm{Si}\left(\frac{t-t_{1}}{100}\right)\right],
\]
where $\mathrm{Si}\left(t\right)=\intop_{0}^{t}\frac{\sin t'}{t'}dt'.$ 

In this way we obtained the following temporal responses at CPA EPs
in the low-Q and high-Q two slab setups in Figs. \ref{fig:time_domain_two_slabs}
and 3 with the parameters in
Figs. \ref{fig:comparison} and 1 (c) and
the temporal response at a CPA for the two-slab setup in Fig. 4
with the parameters above. It can be seen that at CPA EPs the
response to $e^{i\omega_{n}t}$ is without oscillations and the response
to $te^{i\omega_{n}t}$ has an oscillation that decays in time. In
addition, at a standard CPA the response to $e^{i\omega_{n}t}$ has
an oscillation that decays in time and it converts $te^{i\omega_{n}t}$
to $e^{i\omega_{n}t}$ as expected.
\begin{center}
\begin{figure}
\subfloat[]{\begin{centering}
\includegraphics[width=9cm]{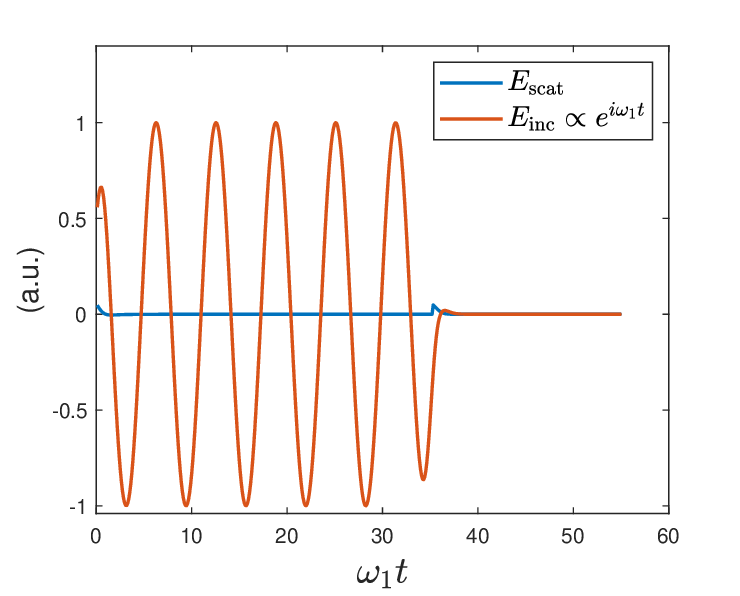}
\par\end{centering}
}\subfloat[]{\begin{centering}
\includegraphics[width=9cm]{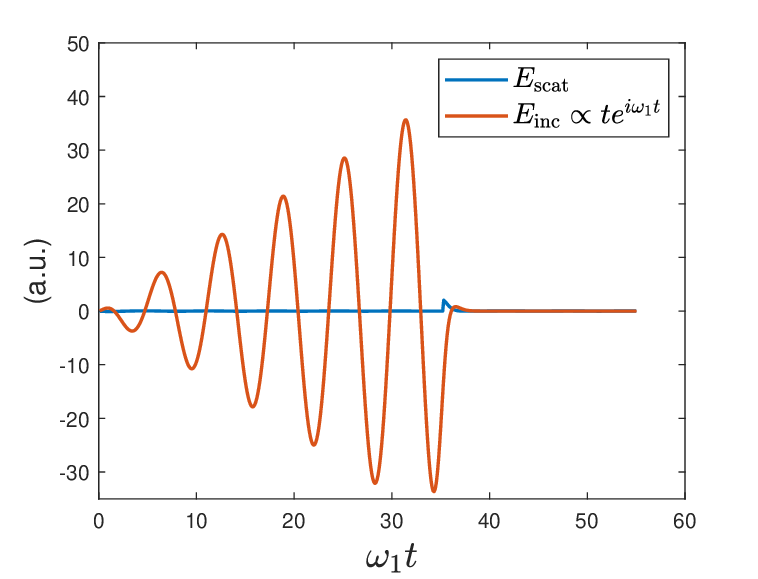}
\par\end{centering}
}

\caption{Scattered wave at a real CPA EP for the inputs of $e^{i\omega_{n}t}$
(a) and $te^{i\omega_{n}t}$ (b) multiplied by $\left(\tanh\left(t\right)+1-\left(\tanh\left(t-40\right)+1\right)\right)/2,$
where $n_{1}=1.295+0.0557i,n_{2}=1.487+0.51i,\,\,l_{1}=1,l_{2}=3.885,\,\,k_{1}=0.88095.$
It can be seen that both inputs are absorbed in CW. \label{fig:time_domain_two_slabs}}
\end{figure}
\par\end{center}

\subsection{Reflection for the input $te^{i\omega t}$}

Multiplying the input by $t$ or an increasing exponent in the input
(virtual CPA/ CPA EP) has an effect that is similar to lowering the
Q of the cavity since the multiple-reflection signal from earlier
times has a low magnitude similarly to the low reflection that causes
the components that originated from the input at earlier time to be
weak. In other words $r\left(t\right)$ and the instantaneous reflected
energy are smaller compared to the standard case of the input $e^{i\omega t}.$ In addition to the output of the form $te^{i\omega t}$, there is the output of $e^{i\omega t},$ which is negligible at large times.

Advantage: lower instantaneous reflection $r\left(t\right)$ and reflected
energy $E_{r}\left(t\right).$

This can be important in applications where the energy delivery until
switching off the signal is important such as ablation.

\section{Conversion between $t^{2}e^{i\omega t}$ and $te^{i\omega t}$ }

We now check the conversion of $t^{2}e^{i\omega t}$ to $te^{i\omega t}$,
which is satisfied when $H^{\left(2\right)}\left(\omega_{n}\right)=H\left(\omega_{n}\right)=0,H'\left(\omega_{n}\right)\neq0.$ 

For the following transfer function $H\left(\omega\right)=\omega-\omega_{n}$
we obtain
\[
H\left(\omega_{n}\right)=0,\,\,\,H'\left(\omega\right)=1,\,\,\,H''\left(\omega\right)=0,
\]
and the conditions above are satisfied.

For a transfer function with a denominator defined as
\[
H\left(\omega\right)=fg,\,\,\,g=\frac{1}{h},
\]
we write
\[
H\left(\omega_{n}\right)=fg=0\Rightarrow f=0,
\]
\[
H'\left(\omega\right)=f'g+g'f=f'g\neq0\Rightarrow f'\neq0,
\]
\[
H''\left(\omega\right)=f''g+f'g'+g'f'+g''f=f''g+2f'g'.
\]
It can be seen that if $f''=0$ e.g., $f=\omega-\omega_{n}$ we have
\[
H''\left(\omega\right)=2f'g',
\]
which we require to vanish. This means that
\[
g'=\left(\frac{1}{h}\right)^{'}=-\frac{h'}{h^{2}}=0
\]
that is satisfied when
\[
h'\left(\omega\right)=0,
\]
as in $h\left(\omega\right)=\left(\omega-\omega_{n}\right)^{2}-C,$ where
$C$ is constant and $h\left(\omega_{n}\right)\neq0$. 

In conclusion, we can have the conversion $t^{2}e^{i\omega t}\rightarrow te^{i\omega t}$
if at $\omega_{n}$
\[
f=0,\,\,f'\neq0,\,\,f''=0,\,\,\,h\neq0,\,\,\,h'=0.
\]

Here there are two conditions $f=0,\,\,h'=0$ and the others are usually
also satisfied. In terms of the degrees of freedom required to satisfy
the conditions it is similar to a CPA EP and we therefore employ a
two-slab setup.

We impose the two conditions for
\[
r=-\frac{r_{1}\left(e^{2ikn_{2}l_{2}}r_{2}+1\right)+e^{2ikn_{1}l_{1}}\left(r_{2}+e^{2ikn_{2}l_{2}}\right)}{1+e^{2ikn_{2}l_{2}}r_{2}+e^{2ikn_{1}l_{1}}\left(r_{2}+e^{2ikn_{2}l_{2}}\right)r_{1}}
\]
and get
\[
r_{1}\left(e^{2ikn_{2}l_{2}}r_{2}+1\right)+e^{2ikn_{1}l_{1}}\left(r_{2}+e^{2ikn_{2}l_{2}}\right)=0,
\]
\[
n_{2}l_{2}e^{2ikn_{2}l_{2}}r_{2}+e^{2ikn_{1}l_{1}}\left(n_{1}l_{1}r_{2}+\left(n_{1}l_{1}+n_{2}l_{2}\right)e^{2ikn_{2}l_{2}}\right)r_{1}=0.
\]

We substitute $x,y$ for the exponents in the first equation and obtain
\[
r_{1}\left(xr_{2}+1\right)+y\left(r_{2}+x\right)=0,
\]
\[
-\frac{r_{1}\left(xr_{2}+1\right)}{r_{2}+x}=y.
\]
We then substitute $y$ in the second equation to get 
\[
n_{2}l_{2}xr_{2}+y\left(n_{1}l_{1}r_{2}+\left(n_{1}l_{1}+n_{2}l_{2}\right)x\right)r_{1}=0,
\]
\[
n_{2}l_{2}xr_{2}-\frac{r_{1}\left(xr_{2}+1\right)}{r_{2}+x}\left(n_{1}l_{1}r_{2}+\left(n_{1}l_{1}+n_{2}l_{2}\right)x\right)r_{1}=0,
\]
\[
n_{2}l_{2}xr_{2}\left(r_{2}+x\right)-r_{1}\left(xr_{2}+1\right)\left(n_{1}l_{1}r_{2}+\left(n_{1}l_{1}+n_{2}l_{2}\right)x\right)r_{1}=0,
\]
\[
n_{2}l_{2}xr_{2}^{2}+n_{2}l_{2}x^{2}r_{2}-r_{1}\left(xr_{2}n_{1}l_{1}r_{2}+xr_{2}\left(n_{1}l_{1}+n_{2}l_{2}\right)x+\left(n_{1}l_{1}r_{2}+\left(n_{1}l_{1}+n_{2}l_{2}\right)x\right)\right)r_{1}=0,
\]
which leads to
\[
x=-\frac{l_{1}n_{1}r_{1}^{2}\left(r_{2}^{2}+1\right)+l_{2}n_{2}(r_{1}-r_{2})(r_{1}+r_{2})\pm\sqrt{l_{1}^{2}n_{1}^{2}r_{1}^{4}\left(r_{2}^{2}-1\right)^{2}-2l_{1}l_{2}n_{1}n_{2}r_{1}^{2}\left(r_{2}^{2}-1\right)\left(r_{1}^{2}+r_{2}^{2}\right)+\text{\ensuremath{l_{2}}}^{2}n_{2}^{2}\left(r_{1}^{2}-r_{2}^{2}\right)^{2}}}{2r_{2}\left(\text{\ensuremath{l_{1}}}n_{1}r_{1}^{2}+l_{2}n_{2}\left(r_{1}^{2}-1\right)\right)},
\]
\[
y=\frac{l_{1}n_{1}r_{1}^{2}\left(r_{2}^{2}-1\right)-l_{2}n_{2}\left(r_{1}^{2}+r_{2}^{2}\right)\pm\sqrt{\left(l_{1}n_{1}r_{1}^{2}\left(r_{2}^{2}+1\right)+l_{2}n_{2}(r_{1}-r_{2})(r_{1}+r_{2})\right)^{2}-4l_{1}n_{1}r_{1}^{2}r_{2}^{2}\left(l_{1}n_{1}r_{1}^{2}+l_{2}n_{2}\left(r_{1}^{2}-1\right)\right)}}{2l_{2}n_{2}r_{1}r_{2}}.
\]
Similarly to the method in the main text in Eq. (4) we obtain the
following solution for two slabs without Bragg mirrors
\[
l_{1}=1,l_{2}=0.25,n_{1}=1.329+0.441i,n_{2}=1.684+1.003i,k=24.76.
\]

\section{The PIC setup parameters}
\begin{figure}[htp]
\begin{centering}
\includegraphics[scale=0.5]{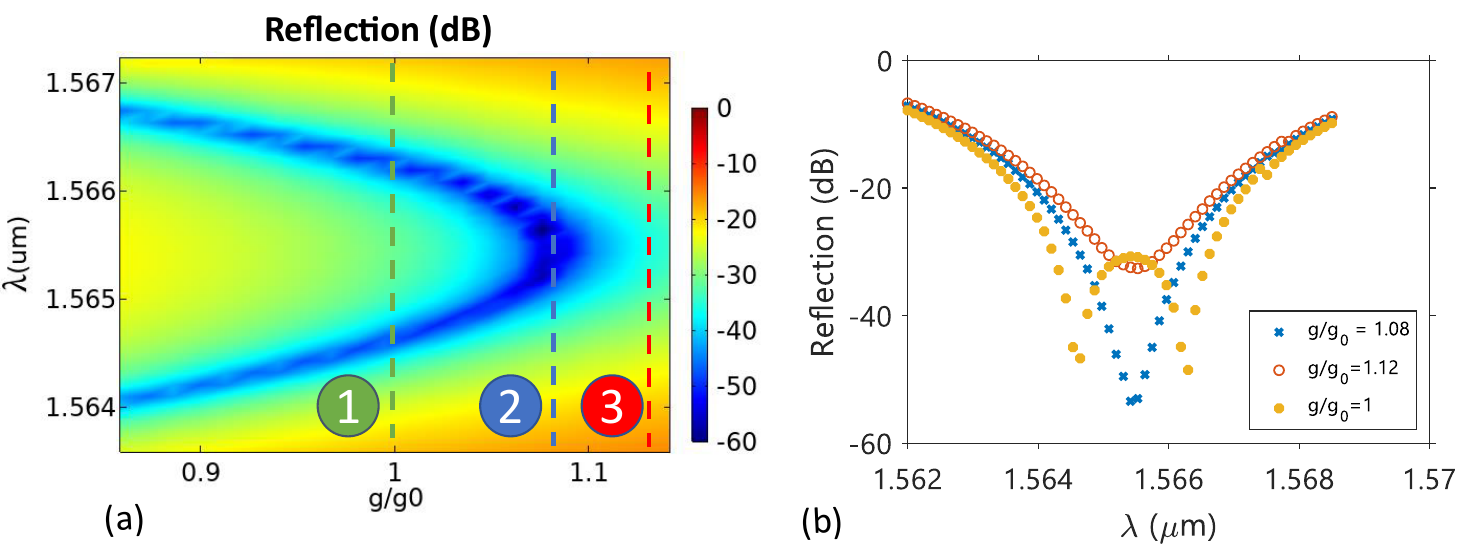}
\par\end{centering}
\caption{(a) Reflection in (dB) scale for varying the geometrical parameter $g/g_0$ of the PIC in Fig. 5 (a). (b) Reflection for different PIC ratios $g/g_0$ as indicated by the dashed lines in (a). }
\label{fig:alu}
\end{figure}

The PIC geometry is based on a dielectric core of Si with a refractive index $n_{\mathrm{Si}}=3.4$ and air cladding, and our simulation domain is considered to be two-dimensional for simplicity. The width of the waveguides and the ring resonator is $300$nm to enable single-mode operation and the radius of the ring  is $3\mu$m in order to have one of its resonances close to $ \lambda_0 = 1.56\mu$m, the optical telecommunication wavelength. The two waveguides are placed parallel to each other and distant from the ring such that the distances of the bottom and top waveguides from the ring are $g_0$ and $g,$ respectively. Additionally, the top waveguide is terminated on both sides with perfect mirrors to allow only two coupling channels to the CROW from the left and/or the right of the bottom waveguide, matching the model in Fig. 1 (a). Here the additional channel indicated by the absorbing boundary in Fig. 5 (a) represents an outgoing mode, and it plays the role of radiation loss to position the CPA EP at real frequencies. Irrespective of the length of the top waveguide, an EP can always be obtained by tuning the coupling ratio between the top and bottom waveguides $g/g_0.$ In general the reflection port indicated in Fig. 5 (a) will show a reflection behavior similar to Eq. (2), where the coupling of the waveguides to the ring and the length of the top waveguide play the role of the reflection at the interfaces and the length of the slabs, respectively. In Fig. \ref{fig:alu} (a) we plot the reflection for various ratios of $g/g_0,$ keeping $g_0=15$nm, which clearly shows that there are two reflection dips at two wavelengths for small values of the ratio $g/g_0$. As this ratio increases, the two dips become closer until they coalesce at  $g/g_0=1.08,$ which corresponds to the CPA EP for $g_{\mathrm{EP}}.$ If we plot the zeros when varying $g/g_0,$ we get a curve analogous to the one shown in Fig. 1 (b), where the two zeros start on the real frequency axis, then coalesce at the EP, and finally emerge into the complex plane, as clearly shown in Fig. \ref{fig:alu} (a) where the two zeros are apparent before the EP and then they fade away after they meet at the EP. It is interesting also to notice that the zeros depart from the EP for lower values of the ratio $g/g_0$ with a square root variation, confirming the EP nature. In addition, in Fig. \ref{fig:alu} (b) we plot the reflection versus wavelength for the three representative data points in Fig. \ref{fig:alu} (a). It can be readily seen that for  $g<g_{\mathrm{EP}}$ there are two reflection dips, then at $g=g_{\mathrm{EP}}$ these two dips meet at $\lambda_{\mathrm{EP}} = 1.5623\mu$m, and finally these two dips disappear and leave only a shallow reflection dip at $\lambda_{\mathrm{EP}},$ indicating that the two zeros are complex with the same real part of the frequency $\mathrm{Re}(\omega) = \omega_{\mathrm{EP}}.$


\bibliographystyle{unsrt}